\documentclass{article}

\usepackage{arxiv}

\usepackage[utf8]{inputenc} 
\usepackage[T1]{fontenc}    
\usepackage{hyperref}       
\usepackage{url}            
\usepackage{booktabs}       
\usepackage{amsfonts}       
\usepackage{nicefrac}       
\usepackage{microtype}      
\usepackage{lipsum}		
\usepackage{graphicx}
\usepackage[numbers]{natbib}
\usepackage{doi}

\usepackage[%
    prefixes-as-symbols=false,
    ]{siunitx}
\usepackage{tikz}
\usepackage{listings}
\usepackage{tikz}
\usetikzlibrary{decorations.pathreplacing}
\usepackage[braket, qm]{qcircuit}
\newcommand{\qlbm}[0]{\textsc{qlbm}}

\definecolor{LightGray}{gray}{0.9}

\makeatletter
\DeclareRobustCommand\onedot{\futurelet\@let@token\@onedot}
\def\@onedot{\ifx\@let@token.\else.\null\fi\xspace}
 
\def\ie{\emph{i.e}\onedot}

\usepackage{listings}

\definecolor{codegreen}{rgb}{0,0.6,0}
\definecolor{codegray}{rgb}{0.5,0.5,0.5}
\definecolor{codepurple}{rgb}{0.58,0,0.82}
\definecolor{backcolour}{rgb}{0.95,0.95,0.92}

\lstdefinestyle{mystyle}{
    backgroundcolor=\color{backcolour},   
    commentstyle=\color{codegreen},
    keywordstyle=\color{magenta},
    numberstyle=\tiny\color{codegray},
    stringstyle=\color{codepurple},
    basicstyle=\ttfamily\footnotesize,
    breakatwhitespace=false,         
    breaklines=true,                 
    captionpos=b,                    
    keepspaces=true,                 
    numbers=left,                    
    numbersep=5pt,                  
    showspaces=false,                
    showstringspaces=false,
    showtabs=false,                  
    tabsize=2,
    frame=tblr
}

\usepackage{cleveref}
\usepackage{lineno}
\usepackage{xcolor}
\usepackage{xspace}
\usepackage{upgreek}
\usepackage{subcaption}
\usepackage{svg}
\usepackage[colorinlistoftodos,prependcaption,textsize=tiny]{todonotes}

\title{\textsc{qlbm} -- A Quantum Lattice Boltzmann Software Framework}

\author{ \href{https://orcid.org/0000-0002-8102-6389}{\includegraphics[scale=0.06]{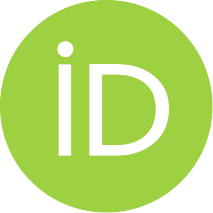}\hspace{1mm}C\u{a}lin A.~Georgescu}\\
	Delft University of Technology\\
	Mekelweg 4, 2628CD, Delft\\
	\texttt{c.a.georgescu@tudelft.nl} \\
	\And
    \href{https://orcid.org/0000-0001-7751-9060}{\includegraphics[scale=0.06]{icons/orcid.pdf}\hspace{1mm}Merel A.~Schalkers}\\
	Delft University of Technology\\
	Mekelweg 4, 2628CD, Delft\\
	\texttt{m.a.schalkers@tudelft.nl} \\
	\And
    \href{https://orcid.org/0000-0003-0802-945X}{\includegraphics[scale=0.06]{icons/orcid.pdf}\hspace{1mm}Matthias M\"{o}ller}\\
	Delft University of Technology\\
	Mekelweg 4, 2628CD, Delft\\
	\texttt{m.moller@tudelft.nl} \\
}

\renewcommand{\shorttitle}{\textsc{qlbm} -- A Quantum Lattice Boltzmann Software Framework}

\hypersetup{
pdftitle={qlbm - A Quantum Lattice Boltzmann Software Framework},
pdfsubject={quant-ph},
pdfauthor={C\u{a}lin A.~Georgescu, Merel A.~Schalkers, Matthias M\"{o}ller},
pdfkeywords={Quantum computing, Lattice Boltzmann Method,
Quantum software},
}

\begin{document}
\maketitle
\begin{abstract}
We present \qlbm, a Python software package designed to
facilitate the development, simulation, and analysis of Quantum Lattice Boltzmann Methods (QBMs).
\qlbm~is a modular framework that introduces a quantum component abstraction hierarchy 
tailored to the implementation of novel QBMs.
The framework interfaces with state-of-the-art quantum software infrastructure
to enable efficient simulation and validation pipelines,
and leverages novel execution and pre-processing techniques
that significantly reduce the computational resources required
to develop quantum circuits.
We demonstrate the versatility of the software by showcasing
multiple QBMs in 2D and 3D with complex boundary conditions,
integrated within automated benchmarking utilities.
Accompanying the source code are extensive test suites,
thorough online documentation resources, 
analysis tools, visualization methods, and demos that aim
to increase the accessibility of QBMs while encouraging reproducibility and collaboration.
The source code of \qlbm~is publicly available
under a permissive MPL 2.0 license at \url{https://github.com/QCFD-Lab/qlbm}.
\end{abstract}


\section{Introduction}

The field of Quantum Computing (QC) \citep{nielsen2010quantum} has received
a staggering amount of attention in recent decades from
researchers and practitioners alike.
Ever since the formulation of the first quantum algorithms in the early 1990s,
quantum computing captured the interest and attention of scientists attempting
to accelerate solvers for high impact, real-life problems.
It was algorithms like those of \citet{deutsch1992rapid},
\citet{bernstein1993quantum},
\citet{grover1996fast}, and
\citet{shor1999polynomial}
that initiated a wave of research aiming to understand
how QC can revolutionize the status quo.
Two properties make the quantum computing paradigm especially attractive
for the increasingly demanding large-scale computational demands of today
-- exponential information compression and quantum parallelism.
The former is a core property of the basic unit of quantum information:
the quantum bit or qubit.
Unlike classical bits, $n$ qubits encode a superposition
that can be represented through a $2^n$-dimensional vector
belonging to a complex Hilbert space.
The latter, quantum parallelism, refers to the ability of
quantum computers to simultaneously encode and update
multiple results in a single computational step.
Thanks to these two properties, QC carries the potential to augment the current 
computational landscape with a drastically different
yet complementary archetype.

The drive to accelerate classical solvers by means of quantum computing
has led to novel quantum algorithms that target nearly every branch of computational science.
From quantum chemistry \citep{o2016scalable, hempel2018quantum, cao2019quantum}
to deep learning \citep{schuld2014quest, jeswal2019recent},
data mining \cite{rebentrost2014quantum, lloyd2014quantum},
and finance \cite{orus2019quantum}, quantum algorithms
promise to augment or improve upon classical methods.
One field where quantum computing advantages are particularly appealing
is that of computational fluid dynamics (CFD).
State-of-the-art CFD simulations are extremely memory- 
and compute-intensive applications that require tremendous
amounts of resources to tackle modern engineering tasks.
It is this computational capacity bottleneck,
together with growing concerns about Moore's law's
\citep{schaller1997moore} future viability that have attracted
many researchers' attention towards the
potentially disruptive effect that
QC could have for CFD simulations \citep{givi2020quantum}.

In recent years, several quantum methods for CFD applications have emerged.
Here, we consider three standout directions for quantum CFD (QCFD) research.
The first two largely center around the Navier-Stokes (NS) governing equations for large-scale, turbulence-minded applications.
Techniques that attempt to directly (approximately)
solve the NS equations with quantum computers are typically
aimed at solving general linear systems of equations (LSE) under specific assumptions.
In particular, the Harrow–Hassidim–Lloyd (HHL) \citep{harrow2009quantum}
algorithm and its subsequent improvements
\citep{ambainis2010variable, clader2013preconditioned, childs2017quantum}
stand out, as they provide theoretical speedups over classical counterparts.
However, in addition to the linearization of the NS equations, 
the viability of the HHL further hinges on
state preparation and amplitude approximation techniques \citep{aaronson2015read}
which may not be practically feasible for CFD applications.

A second way in which quantum computers can solve LSEs
consists of so-called variational quantum algorithms.
Variational Quantum Linear Solvers (VQLS) such as those put forward by
\citet{bravo2023variational} and \citet{patil2022variational}
attempt a task similar to that of HHL, but instead rely on
parameterized quantum circuits, the parameters of which
are iteratively improved.
Recently, research has shown that such techniques could be
used to solve Stokes flow \citep{liu2024variational}
and the heat conduction equation \cite{liu2022application}.
\citet{kyriienko2021solving} also introduce a framework
of differentiable quantum circuits and feature maps-based encodings
and use it solve the quasi-1D NS equations, while
\citet{paine2023quantum} extend the variational quantum algorithm (VQA) framework
to support \emph{kernel methods}, a technique of embedding
data into higher-dimensional spaces \citep{schuld2019quantum},
and demonstrate the ability to solve ordinary differential equations.
An advantage of VQAs is their relatively
shallow circuits, which makes them
suitable candidates for the quantum hardware available today.
However, they too suffer from several important limitations.
Optimizing the parameters of VQAs is a computationally intensive
task, that is delegated to classical computers.
This introduces significant overhead not only in 
the classical optimization procedure of the parameters, but also in the
the quantum-classical communication channel.
Moreover, optimizing VQA circuits might not be computationally feasible,
due to the barren plateau problem
in quantum computing \citep{mcclean2018barren, larocca2024review}.

The third way in which QCFD problems can be approached
is through the quantum implementation of \emph{Lattice Boltzmann Methods} (LBMs).
This avenue presents modelling opportunities for
exploiting the mathematical
structure of the Boltzmann Equation and is entirely independent
from classical optimization requirements.
It is this direction that we seek to advance through this work.
In what follows, we describe the current landscape of quantum LBM research before
highlighting the challenges currently facing this field and how
this work seeks to address them in \Cref{subsec:software-and-lbms}.
\Cref{subsec:lbm} describes the steps and the mathematical structure
of the classical LBM.


Recently, quantum lattice Boltzmann methods (QBMs) have emerged
as promising candidates for the future direction of QCFD.
While the physics that QBMs target 
is entirely classical, the principal premise behind QBMs
is that quantum computing may enable simulation
at scales otherwise unattainable with classical hardware.
The linearity of the streaming step and the locality of collision
are two of the reasons why the LBM lends
itself particularly well to native quantum implementations.
Despite this, there are several inherent caveats that
quantum implementations must address to
simulate physically correct behavior.
Most notably, these include the nonlinearity of the collision operator
and the nonlocality of the streaming operator.
These challenges stem from the fundamental properties
of physical mesoscopic and macroscopic fluids, and are
ubiquitous across many governing equations in science and engineering.
One additional benefit of the LBM is that unlike in
the Navier-Stokes equations, the nonlinearity and nonlocality
are not directly coupled, which provides promising modelling opportunities.
Supplementary to equation-specific nuances,
effectively encoding information into
and extracting it out of the quantum state
are two universal hurdles of quantum algorithms.
In an effort to overcome these challenges,
research surrounding QBMs has largely focused on the development
of quantum primitives that implement (parts of) the LBM time-marching loop.
These initiatives have given rise to several techniques that accommodate
specific subroutines of the LBM, imposing trade-offs between scalability and versatility.
One way to categorize existing QBMs is
by how they address the inherent nonlinearity of collision.

The initial wave of research into QCFD
occurred between 2001 and 2003 and largely focused on
extending the lattice-gas model to distributed quantum devices
\citep{yepez2001quantum, yepez2002efficient, yepez2002quantum, pravia2003experimental}.
This work tailors quantum lattice-gas solvers to
a decentralized system of quantum computers
with limited number of qubits per device, linked together
through classical communication channels.
Though this approach enables the balancing of the computational workload
through horizontal scaling, it requires a number of qubits
that grows linearly with the number of grid points of the lattice.

\citet{todorova2020quantum} and
\citet{schalkers2024efficient}
propose \emph{collisionless} methods that include primitives for particle streaming and
boundary conditions, but omit the collision operator entirely.
\citet{steijl2020quantum} and \citet{moawad2022investigating}
alternatively propose a method in which
quantum primitives that implement floating point arithmetic
can compute nonlinear terms, but require 
a reversible conversion between the encoding
of the quantum state used to perform streaming
and the encoding that enables the computation of the nonlinear velocity terms
at each time step.
\citet{itani2022analysis} and \citet{sanavio2024lattice} 
adopt an approach based on truncated Carleman linearization, that
approximates the non-linear LBE by a finite-dimensional
linear system of equations that can be expressed in terms
of (unitary) quantum operators.
However, these approaches require a large number
of additional variables that detract from scalability
and which do not naturally decompose into quantum circuits.
Budinski \cite{budinski2021quantum, ljubomir2022quantum} further
developed an approach that enables both streaming and collision
but that incurs a probability of measuring an orthogonal 
(irrelevant) quantum state after each time-step.
\citet{frankel2024quantum} propose a similar approach,
which, similar to Budinski's methods,
requires the costly decomposition of unitary 
matrices into quantum gates for compatiblity with quantum hardware.
More recently, \citet{wawrzyniak2024quantum} introduced
a novel method based on the same linear combination
of unitaries  (LCU) \citep{childs2012hamiltonian}
approach tailored to the advection-diffusion equation,
but which also requires full state measurement and reinitialization after each time step.
Finally, \citet{schalkers2024importance} 
extended a previously developed encoding and
equipped it with a collision operator inspired by lattice gas automata
at the cost of requiring a number of qubits that scales
with the number of simulated time steps up to grid size.

The current state of QBMs is fragmented between several
approaches that each present different strengths and weaknesses.
This poses several challenges for researchers seeking to advance the field.
In what follows, we highlight three significant
challenges that face the development of QBMs, draw parallels
to their classical counterparts, and explain how software can help mitigate these issues.

\subsection{Software and QBM Research \label{subsec:software-and-lbms}}

To advance the theory of QBMs researchers require infrastructure
that enables the implementation and experimentation of their algorithms.
We address these concerns by drawing parallels to the more mature classical LBM
field, and the methods that have emerged to facilitate its practical success.
We then discuss the absence of such methods from the QBM field,
and the drawbacks that researchers face because of this.
Finally, we address how the current work seeks to mitigate these shortcomings.

Classical LBMs owe their popularity to several factors.
From a theoretical standpoint, LBMs allow for
the computation of macroscopic quantities
such as mass and momentum density \cite{kruger2017lattice},
and can be used as (approximate) solvers for Navier-Stokes applications,
among other target equations \cite{chai2008novel, wang2011lattice, chai2013lattice}.
From a practical standpoint, the LBM lends itself well to 
massively parallel computing paradigms
\cite{succi2001lattice, kruger2017lattice}.
Over the years, several parallel software implementations of
the LBM have emerged, including \textsc{HemeLB} \cite{mazzeo2008hemelb}
\textsc{openLB} \cite{krause2021openlb}, 
\textsc{Palabos} \cite{latt2021palabos},
\textsc{waLBerla} \cite{bauer2021walberla},
\textsc{lbmpy} \cite{bauer2021lbmpy},
and \textsc{pylbm} \cite{pylbm},
which are able to carry out distributed simulations
on hundreds of heterogeneous compute nodes.
In addition to practical applications, open-source LBM
software implementations have another significant merit --
they facilitate the development of further research by establishing
a foundation for both theory and infrastructure \cite{krause2021openlb}.

Such foundations are almost entirely absent in the realm of QBMs.
Because of this, the field faces three distinct hurdles.
First, the many nuances of present QBM techniques
make the comparison of the performance and scalability difficult.
From various quantum state encodings \cite{schalkers2024importance} to the decomposition
of exponentially sized matrices into quantum gates, QBMs
build on top of extensive knowledge and technology stacks
that make implementation a daunting challenge.
Second, the fractured nature of the field poses
challenges for techniques that augment existing work,
such as the effective extraction of quantities
of interest from the quantum state \cite{schalkers2024momentum}.
Third, due to the scarce availability of QBM implementations,
researchers face the additional obstacle of
verifying and comparing methods from the literature.
This significantly detracts from the reproducibility of the field.
Before addressing how software can help ameliorate these three challenges,
we first introduce the current state of \emph{quantum software}
and its relation to present day quantum computer hardware.

The current state of QC hardware has been undergoing rapid
development and is currently in the so-called 
\emph{Noisy Intermediate-Scale Quantum} phase \cite{preskill2018quantum}.
While quantum computers available today showcase some of the
core advantages that theoretical physics promises, they
are limited in both the number of qubits available
and the time span that qubits can retain coherent states.
These constraints greatly impede on the
applications that quantum computers can presently carry out.
To facilitate the research of quantum algorithms 
in an era without \emph{Fault-Tolerant} Quantum Computers,
scientists have turned to simulation methods instead.
Recently, an increasing number software frameworks have emerged
to bridge the gap between theoretical advances in algorithmics
and hardware availability.
These range from general purpose simulation tools
\cite{qiskit2024, suzuki2021qulacs, sivarajah2020t}
to specialized packages aimed at machine learning 
\cite{bergholm2018pennylane, broughton2020tensorflow} 
and material simulation \cite{bassman2022arqtic}. 
The current state of \emph{quantum software} intersects
QC theory and available hardware, such that
researchers can leverage classical hardware to verify large-scale algorithmic 
prototypes while quantum counterparts edge closer to fault-tolerance.

In this work, we seek to address the three challenges facing QBM research
by introducing the \qlbm~software framework.
With \qlbm, we aim to bring the same advantages
that classical LBM software has proven to offer to researchers and practitioners alike.
We design the \qlbm~software around the current paradigm
of simulation, with the goal of accelerating QCFD research
in the absence of fault-tolerant quantum computers.
Achieving this requires addressing several challenges, including 
integration with available software and hardware infrastructure,
establishing suitable data structures and design patterns for the development of QBMs,
and providing this functionality in a package that is flexible enough
to conduct research, yet accessible enough for new users.
To the best of our knowledge, \qlbm~is only the second
effort to generalize the software development process of QBMs.
Recently, \citet{shinde2024utilizing} introduced a software tool
aimed at developing and simulating QBMs using the Intel Quantum SDK
and quantum hardware.
However, their work focuses on hybrid
quantum-classical QBM algorithms such as \cite{budinski2021quantum} and \cite{ljubomir2022quantum}
and is specifically targeted towards a single vendor, and not openly available.
By contrast, \qlbm~focuses on fully quantum approaches,
provides a more flexible set of tools from multiple vendors,
is readily available on GitHub under a permissive license,
and pursues the broader goal of providing
an end-to-end development environment.
We describe the internal design of
\qlbm~and the simulations it enables in \Cref{sec:overview}.
\Cref{sec:results} provides results that showcase the
capabilities of \qlbm, both in terms of the QBMs
that it can simulate, as well as the performance improvements it provides.

\subsection{Lattice Boltzmann Methods \label{subsec:lbm}}

To provide additional background for QBMs,
this section briefly introduces the classical
formulation of the Lattice Boltzmann Method (LBM)
and components.
For a more wholistic overview of the LBM,
we refer the reader to the works of
\citet{succi2010lattice} and \citet{kruger2017lattice}.
The Boltzmann Equation (BE) describes the kinetic behavior of fluid
at the mesoscopic scale, nestled between microscopic Newtonian dynamics
and macroscopic Navier-Stokes continua.
The BE models the state of populations of fluid particles
as a statistical distribution function over
physical space, velocity, and time.
\Cref{eq:be} gives the form of the BE we consider throughout this work,
where the left hand-side terms
model the advection of particles over the phase space, and the $\Omega(f)$ term
represents the change in state as a result of particle collisions, often referred
to as the \emph{collision operator}.

\begin{equation}
    \frac{\partial f}{\partial t} + \mathbf{u}\frac{\partial f}{\partial \mathbf{x}} = \Omega(f) 
    \label{eq:be}
\end{equation}

Though several collision operators have been developed over the decades,
the Bhatnagar-Gross-Kook (BGK) formulation
\cite{bhatnagar1954model} remains one of the more
popular and widely implemented options thanks to 
its theoretical and computational simplicity
and its ability to recover the same bulk
properties as Navier-Stokes simulations \cite{kruger2017lattice}.
The BGK collision operator is defined by $\Omega(f) = -\frac{1}{\tau}(f - f^{eq})$
and models the relaxation of the particle distribution $f$ towards the equilibrium
function $f^{eq}$,
where $\tau$ is referred to as the \emph{relaxation time}, and directly
influences the computation of transport coefficients.
The discretization of the BE along phase space and time yields
the Lattice Boltzmann Equation (LBE), which can in turn be solved
numerically by the Lattice Boltzmann Method.
Equipping the BE with the BGK collision operator and discretizing in
terms of physical space, velocity space, and time
yields the most widely-used
form of the LBE \cite{kruger2017lattice}, as described in \Cref{eq:lbe}.

\begin{equation}
    f_i(\mathbf{x} + \mathbf{v}_i \Delta t, t + \Delta t) = f_i(\mathbf{x}, t) - \frac{\Delta t}{\tau} (f_i(\mathbf{x}, t) - f_i^{eq}(\mathbf{x}, t))
    \label{eq:lbe}
\end{equation}

The subscript $i$ of the $f_i$ and $v_i$ variables stems from the 
velocity space discretization, which spans a small set
of discrete velocity channels that particles can travel across.
Including the velocity variable in the subscript rather than a parameter to the
function is a notational convention.
The $f_i$ terms are terms referred to as particle
\emph{populations}, and $\mathbf{v}_i$ terms model velocity
coefficient vectors according to the discretization scheme.
\Cref{eq:eq-function} gives typical choice of equilibrium function for Navier-Stokes simulations
of isothermal models, where $c_s = \Delta x / \Delta t$ is the lattice speed,
$w_i$ are pre-determined weights, $\rho$ is the fluid density,
and $\mathbf{u}$ corresponds to the flow velocity.

\begin{equation}
    f_i^{eq}(\mathbf{x}, t) = w_i \rho \left(1 + \frac{\mathbf{u} \cdot \mathbf{v}_i}{c_s^2} + \frac{(\mathbf{u} \cdot \mathbf{v}_i)^2}{2c_s^4} - \frac{\mathbf{u} \cdot \mathbf{u} }{2c_s^2} \right)
    \label{eq:eq-function}
\end{equation}

The LBM describes a class of time step
algorithms that iterate through repeated steps.
Each time step can be conceptually broken down into three subroutines:
streaming through physical space,
reflection at the boundaries of the fluid domain,
and (non-linear) particle collision.
First, the particles \emph{stream} (or \emph{propagate}) through space in
directions prescribed by the discretized
velocities to neighbouring lattice points.
Then, boundary conditions are applied to
ensure particles adhere to the fluid domain.
Finally, the populations undergo collision (or relaxation) 
before the computation of macroscopic forces.
In addition to Navier-Stokes applications, the different building
blocks of the LBM lend themselves well to a broad range of other use cases.
Among others, researchers have proposed LBM-based models
for acoustics \cite{buick1998lattice}, 
electro-osmotic flows \cite{guo2005lattice},
as well as the Poisson \cite{chai2008novel},
shallower water \cite{zhou2004lattice},
and advection-diffusion \cite{chai2013lattice} equations.
\section{\qlbm~Overview\label{sec:overview}}

\begin{figure}
	\centering
	\input{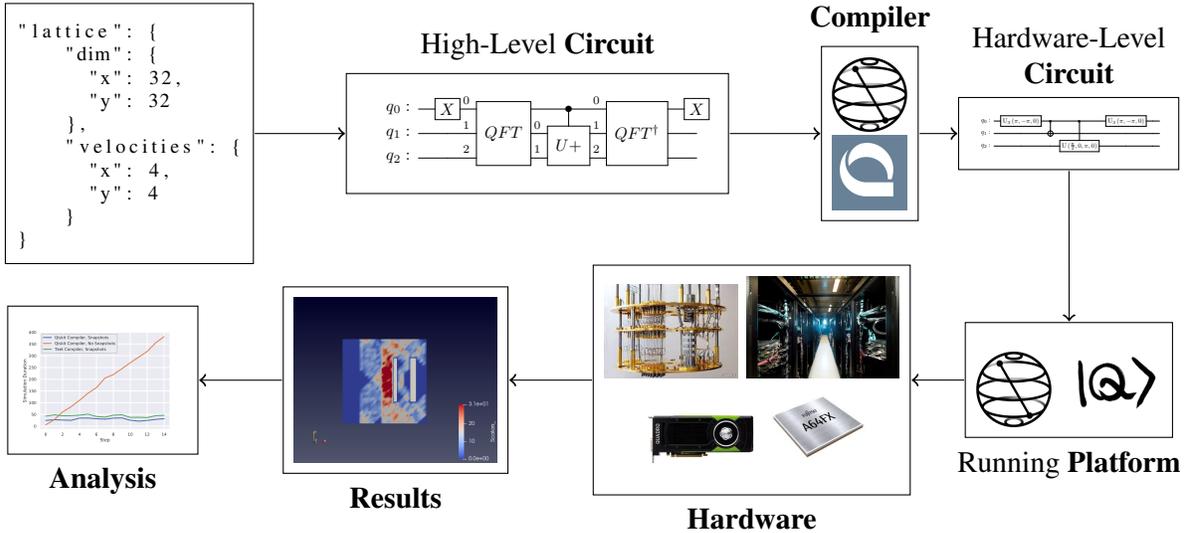}
	\caption{Overview of end-to-end \qlbm~workflow. }
	\label{fig:qlbm-workflow}
\end{figure}

This section introduces the cornerstone features of \qlbm. 
Before addressing these features, however, we
first highlight the end-to-end workflow we designed \qlbm~around.
The primary goal of \qlbm~is to provide an end-to-end environment
for the development, simulation, and analysis of QBM algorithms.
\Cref{fig:qlbm-workflow} provides a visualization of
the workflow that accommodates all of these steps.
The process can be broadly broken up into three substeps:
quantum circuit generation, simulation, and analysis.
We design the \qlbm~workflow as a multi-step pipeline,
where the output of each step is seamlessly forwarded to the next,
while retaining individual access points for user analysis and intervention.
The current implementation of the \qlbm~pipeline supports two algorithms: 
the Quantum Transport Method (QTM) \citep{schalkers2024efficient}
and the Space-Time Quantum Boltzmann Method (STQBM) \citep{schalkers2024importance}.

The workflow begins with a user-friendly specification
of the system to simulate.
The goal of this interface to increase the accessibility
of QBM algorithms for users with limited experience in the field,
while simultaneously enriching the experience of more mature practitioners.
User-specified data includes
information about the lattice discretization,
as well as geometry and boundary conditions.
\qlbm~parses this configuration and extracts algorithm-specific
information that is then used to generate high-level quantum circuits.
This method of deriving circuit properties from high-level
specification bridges the gap between the expectations of
end-users who are looking to perform CFD simulations
and the complexity of specifying physically accurate
quantum algorithms.
We address the internal design choices that facilitate this process
in more detail in \Cref{subsec:internal-architecture}.

Once the high-level quantum circuit has been assembled,
users are generally interested in simulating the algorithm
to verify its correctness and to analyze results.
To make the best use of available resources,
software should exploit  techniques that quantum simulators allow for that
would otherwise not be available on quantum hardware.
To this end, \qlbm~implements methods that lessen the computational burden
on both the algorithmic and the computational fronts.
We describe such techniques and how \qlbm~leverages them to
exploit the time-marching nature of
LBMs in more detail in \Cref{subsec:performance-enhancements}.

Finally, after simulations have concluded, researchers
are typically interested in the performance and scalability
of the methods they are developing.
To accommodate this need, we integrate \qlbm~with a set of tools
that enable the analysis of quantum circuits and their performance.
These tools include means for visualizing QBM algorithms
and their building blocks, exporting simulation results
to external visualization engines, and scripts that
give insight into the scalability of the methods.
We delve into more details on how
\qlbm~integrates with surrounding
quantum software infrastructure in \Cref{subsec:interfacing-integration}.

\subsection{Internal Architecture \label{subsec:internal-architecture}}

The internal architecture of \qlbm~primarily targets two 
goals with regard to quantum circuits.
First, the quantum circuit components of \qlbm~should
be easy to extend and verify,
as to facilitate the design of novel QBM algorithms.
Second, the internal composition of the framework should be
modular, as to enable testability on individual methods through isolation.
In addition to quantum circuit design, the architecture of the
software should minimize the effort required to integrate with external
quantum software libraries.

We address the two design directions of \qlbm~-- quantum components
and overall system design -- in \Cref{subsubsec:qc-arch}
and \Cref{subsubsec:system-arch} respectively, zooming out from
individual quantum circuit abstractions to a holistic overview of the framework.

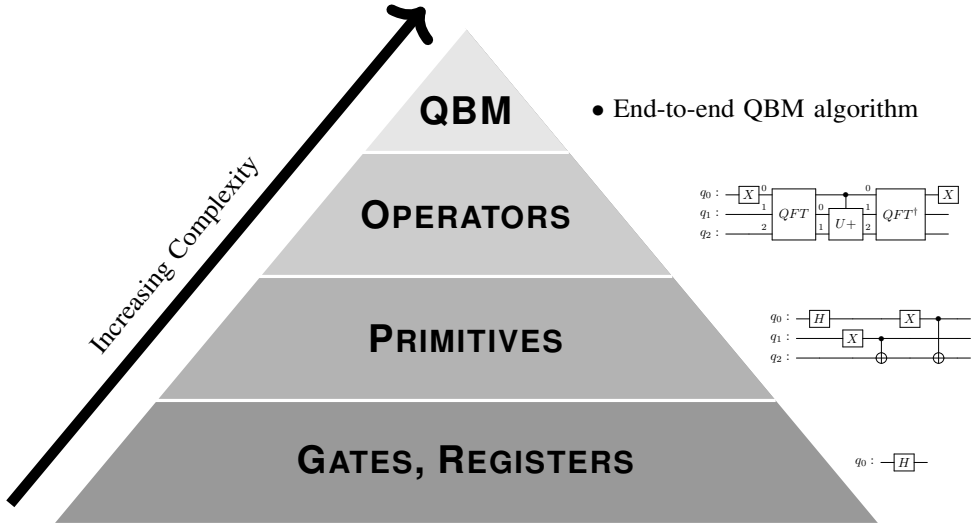
\begin{figure}
	\centering
	\begin{tikzpicture}[scale=0.55,align=center]
\usetikzlibrary{positioning,shapes.arrows,arrows.meta,shadows,backgrounds,fit}

        \filldraw[very thick,white,fill=gray!80] (0,0) -- (-10,-12) -- (10,-12);
        \filldraw[very thick,white,fill=gray!60] (0,0) -- (-7.5,-9) -- (7.5,-9);
        \filldraw[very thick,white,fill=gray!40] (0,0) -- (-5,-6) -- (5,-6);
        \filldraw[very thick,white,fill=gray!20] (0,0) -- (-2.5,-3) -- (2.5,-3);        
        \node at (0,-2) {\bfseries\sffamily\scshape\Large QBM};
        \node at (0,-4.5) {\bfseries\sffamily\scshape\Large Operators};
        \node at (0,-7.5) {\bfseries\sffamily\scshape\Large Primitives};
        \node at (0,-10.5) {\bfseries\sffamily\scshape\Large Gates, Registers};
        \node[text width=8cm] at (9.5,-7.5) {\scalebox{0.5}{
\Qcircuit @C=1.0em @R=0.2em @!R { \\
	 	\nghost{{q}_{0} :  } & \lstick{{q}_{0} :  } & \gate{{H}} & \qw & \qw & \gate{{X}} & \ctrl{2} & \qw & \qw\\
	 	\nghost{{q}_{1} :  } & \lstick{{q}_{1} :  } & \qw & \gate{{X}} & \ctrl{1} & \qw & \qw & \qw & \qw\\
	 	\nghost{{q}_{2} :  } & \lstick{{q}_{2} :  } & \qw & \qw & \targ & \qw & \targ & \qw & \qw\\
\\ }}};
        \node[text width=8cm] at (10,-10.5) {\scalebox{0.5}{
\Qcircuit @C=1.0em @R=0.2em @!R { \\
	 	\nghost{{q}_{0} :  } & \lstick{{q}_{0} :  } & \gate{{H}} & \qw\\
\\ }}};
        
        \begin{scope}[xshift=-1cm,yshift=0.5cm]
            \draw[->, line width=4pt, black] (-10,-12) -- (0, 0);
            \node[text width=10cm,rotate=50.5] at (-6,-6) {Increasing Complexity};
        \end{scope}

        \node[text width=8cm] at (8.5,-4.5) { \scalebox{0.5}{
\Qcircuit @C=1.0em @R=0.2em @!R { \\
	 	\nghost{{q}_{0} :  } & \lstick{{q}_{0} :  } & \gate{{X}} & \multigate{2}{{QFT}}_<<<{0} & \ctrl{1} & \multigate{2}{{QFT^\dagger}}_<<<{0} & \gate{{X}}\\
	 	\nghost{{q}_{1} :  } & \lstick{{q}_{1} :  } & \qw & \ghost{{QFT}}_<<<{1} & \multigate{1}{{U+}}_<<<{0} & \ghost{{QFT^\dagger}}_<<<{1} & \qw\\
	 	\nghost{{q}_{2} :  } & \lstick{{q}_{2} :  } & \qw & \ghost{{QFT}}_<<<{2} & \ghost{{U+}}_<<<{1} & \ghost{{QFT^\dagger}}_<<<{2} & \qw\\
\\}}};
        
        \node[text width=8cm] at (7,-2) {$\bullet$ End-to-end QBM algorithm};
\end{tikzpicture}
	\caption{Representation of internal quantum circuit abstraction hierarchy.}
	\label{fig:component-pyramid}
\end{figure}

\subsubsection{Quantum Component Architecture}
\label{subsubsec:qc-arch}

To realize a modular and extendable quantum circuit library, 
we implement a system of circuit abstractions
based around a complexity hierarchy with respect to the steps of the LBM.
\Cref{fig:component-pyramid} provides a graphical depiction of this hierarchy.
We organize components in three broad categories:
primitives, operators, and algorithms.
Primitives are the least complex elements of
the taxonomy, and they implement small-scale,
isolated blocks within QBM algorithms.
Isolating parameterized implementations of such circuits
enables developers to verify their behavior in isolation
and reuse them seamlessly.
This in turn accelerates the implementation of novel algorithms.

\begin{figure}
    \centering
    \hfill
    \subcaptionbox{Standard controlled incrementer circuit.\label{fig:incrementer-standard}}{\input{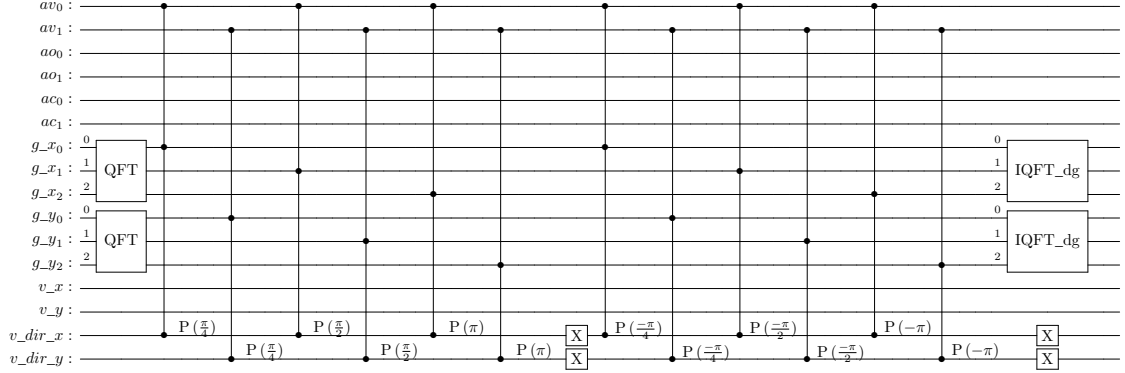}}%
    \hfill%
    \subcaptionbox{Bounceback reflection controlled incrementer.\label{fig:incrementer-bounceback}}{\input{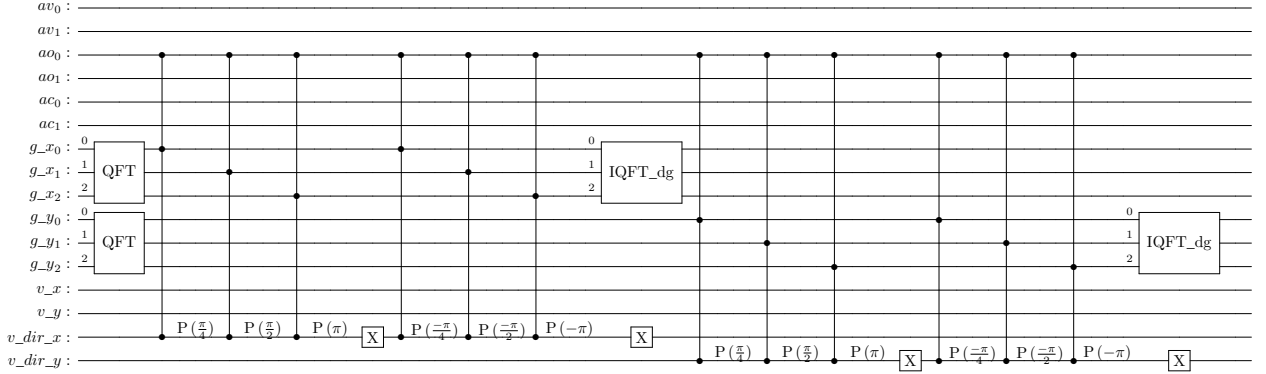}}%
    \caption{Comparison of \qlbm~primitive quantum circuits.}
    \label{fig:controlled-incrementer}
\end{figure}

\Cref{fig:controlled-incrementer} contains an example of how
programmatic quantum circuit construction helps simplify the
algorithmic development process.
Both circuits in the figure were constructed with simple calls to
a \texttt{ControlledIncrementer} primitive that is used
repeatedly to move particles during the streaming
and reflection steps of the QTM algorithm.
The circuits share the same structure: they begin by
mapping the grid qubits to the Fourier basis by performing
a Quantum Fourier Transform (QFT)
and conclude by returning them to the computational basis.
Between the two QFT blocks is a series of controlled phase shifts
that performs the incrementation of the appropriate populations.
In \Cref{fig:incrementer-standard}, the controls
reside on the ancilla velocity qubits, which determine whether
particles move within one CFL substep.
In \Cref{fig:incrementer-bounceback}, the phase shift
controls are instead placed on the ancillary qubit that determine whether
particles have virtually streamed inside of an obstacle.
This sole discrepancy determines which populations are streamed
and differentiates two distinct phases of the algorithm.
The \qlbm~implementation of this primitive allows
the same piece of code to construct both circuits with a single
parameter switch between \texttt{reflection=None} and \texttt{reflection="bounceback"}.

A step above primitives are so-called operators.
The goal of operators is to encompass quantum circuits
that implement one specific physical operation of the LBM --
streaming, reflection, or collision.
This layer of abstractions seeks to address the fragmented
formulations that have emerged from recent QBM literature.
For instance, circuits that perform streaming in
basis-state encodings are not directly applicable
to amplitude-based encodings, and vice-versa.
Despite this fundamental incompatibility,
a clear separation between operator-level circuits 
in different encodings serves two purposes.
First, operators enable targeted experimentation with competing
implementations such as different
boundary condition circuits within a single encoding.
Second, this design allows for a broader
range of algorithmic combinations, which 
may target different solvers or equations.

The highest level of abstraction within the 
component taxonomy is the end-to-end QBM algorithm.
These structures are crucial for tying together
lower level abstractions.
Algorithm-level components follow naturally from the
chaining of operators in a way that resembles LBM pseudocode.
\Cref{fig:cqbm-code} depicts an example of how
the QTM algorithm \cite{schalkers2024efficient} can be expressed
as a series of four operator-level components.
The \texttt{CollisionlessStreamingOperator} first prepares
the correct ancilla qubit state based on the current state
of a CFL counter, before performing a controlled incrementation
on populations of particles within one substep (line 5).
Following streaming, the \texttt{SpecularReflectionOperator} (line 12)
and \texttt{BounceBackReflectionOperator} (line 19)
append the circuit with logic that detects whether particles have
streamed outside the fluid domain, before inverting the appropriate velocities and
placing the populations back in the fluid domain.
Finally, the \texttt{StreamingAncillaPreparation} operator (line 27)
prepares the quantum state for the next iteration of the CFL counter.
A simple call to the built-in class, \ie \texttt{CQLBM(lattice)}
\footnote{In the software, we denote algorithms by their
interpretation of the LBM. For this reason, the QTM algorithm \cite{schalkers2024efficient}
is available as \texttt{CQLBM}, short for \emph{Collisionless} Quantum LBM,
as it models $\Omega(f) = 0$. 
The STQBM \cite{schalkers2024importance} is available as \texttt{SpaceTimeQLBM}.},
is all that users need to do to build end-to-end quantum circuits.

\begin{figure}
	\centering
	\input{code/cqbm}
	\caption{Sample \qlbm~operator code.}
	\label{fig:cqbm-code}
\end{figure}

This internal architecture of quantum components
enables the development of new QBM circuits in two ways.
First, the quantum circuits already implemented
within \qlbm~are trivially reusable
for new methods, provided that encodings are compatible.
Second, this hierarchy facilitates the development
of entirely novel algorithms by additionally separating
quantum circuit logic and quantum register setup.
We achieve this by isolating the quantum register logic
within implementations of the \texttt{Lattice} class,
which are algorithm- and implementation-specific.
Consider again the chaining of operators depicted in \Cref{fig:cqbm-code}.
The only information required to construct the quantum operators
already resides in the \texttt{lattice} attribute of the
\texttt{CQLBM} object, which gets propagated
down the abstraction chain.
To increase the accessibility of this architecture,
we additionally provide each \texttt{Lattice} class
with methods that allow for human-readable
indexing operations by assigning each register
an intuitive naming scheme, and automatically adjusting its size.
This alleviates the burden of manually indexing individual qubits
and addressing multiple logically connected indices.
In addition to the QTM 
algorithm, \qlbm~also fully supports
the STQBM \cite{schalkers2024importance},
which uses a different, extended computational basis state encoding,
which highlights the versatility of this design.

\begin{figure}
	\centering
	\includegraphics[scale=0.75]{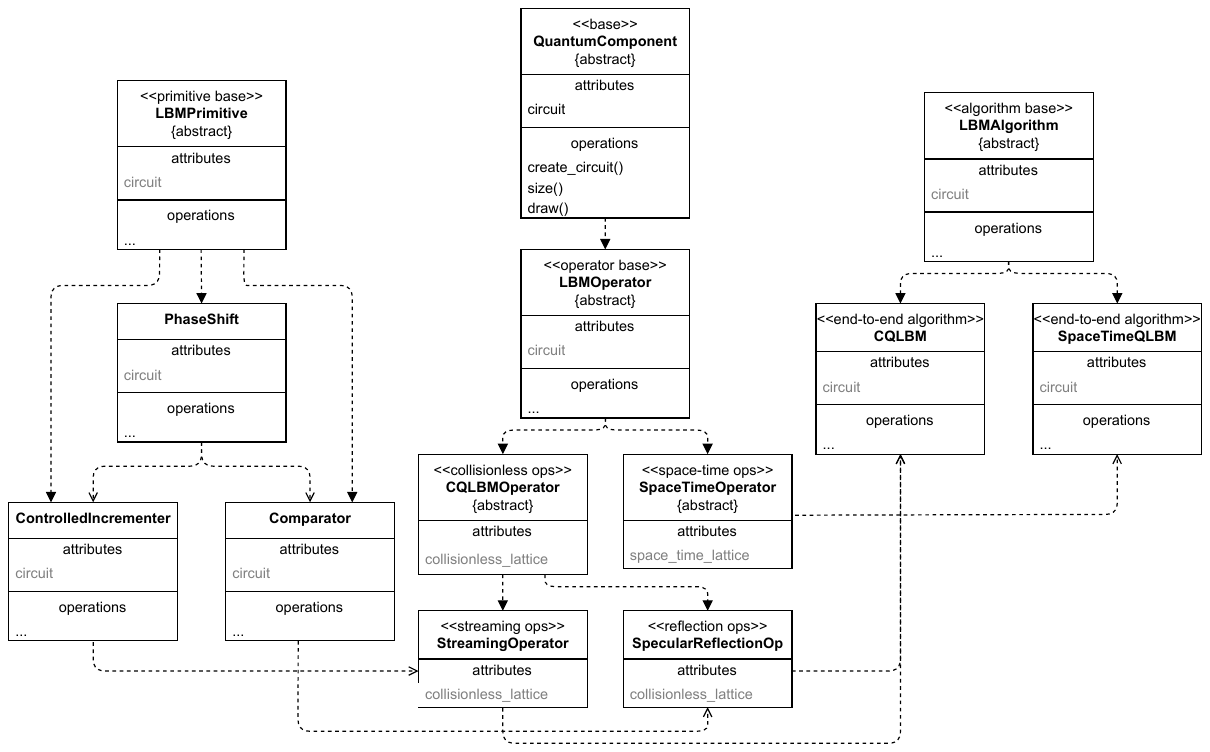}
	\vspace{-11cm}
	\caption{Class diagram representation of the \qlbm~quantum  component architecture.}
	\label{fig:class-diagram}
\end{figure}

\Cref{fig:class-diagram} depicts the  entire architecture of the quantum
components of \qlbm.
At the top, three base classes that 
adhere to the primitive, operator, QBM model
provide interfaces that ease
the development of novel circuits by providing
appropriate interfaces through inheritance.
On the vertical axis, classes become increasingly specific
and complex from top to bottom.
Within one "branch" of the inheritance hierarchy,
component reuse is still possible, \ie by
utilizing simpler primitives to build more complex ones.
On the horizontal axis, components again range from simple to complex
with respect with the task they fulfill within the QBM.
That is, incrementers and comparators serve as the building blocks
for streaming and reflection operators, which then assemble the end-to-end QBM.
Researchers can develop novel QBMs along this axis,
in parallel to existing implementations.
In practice, this leads to a system in which previous contributions
are available for novel developments, but do not hinder them.
The following subsection describes how quantum component module
fits within the broader scope of the framework.

\subsubsection{System Architecture}
\label{subsubsec:system-arch}

\begin{figure}
	\centering
	\includegraphics[scale=0.7]{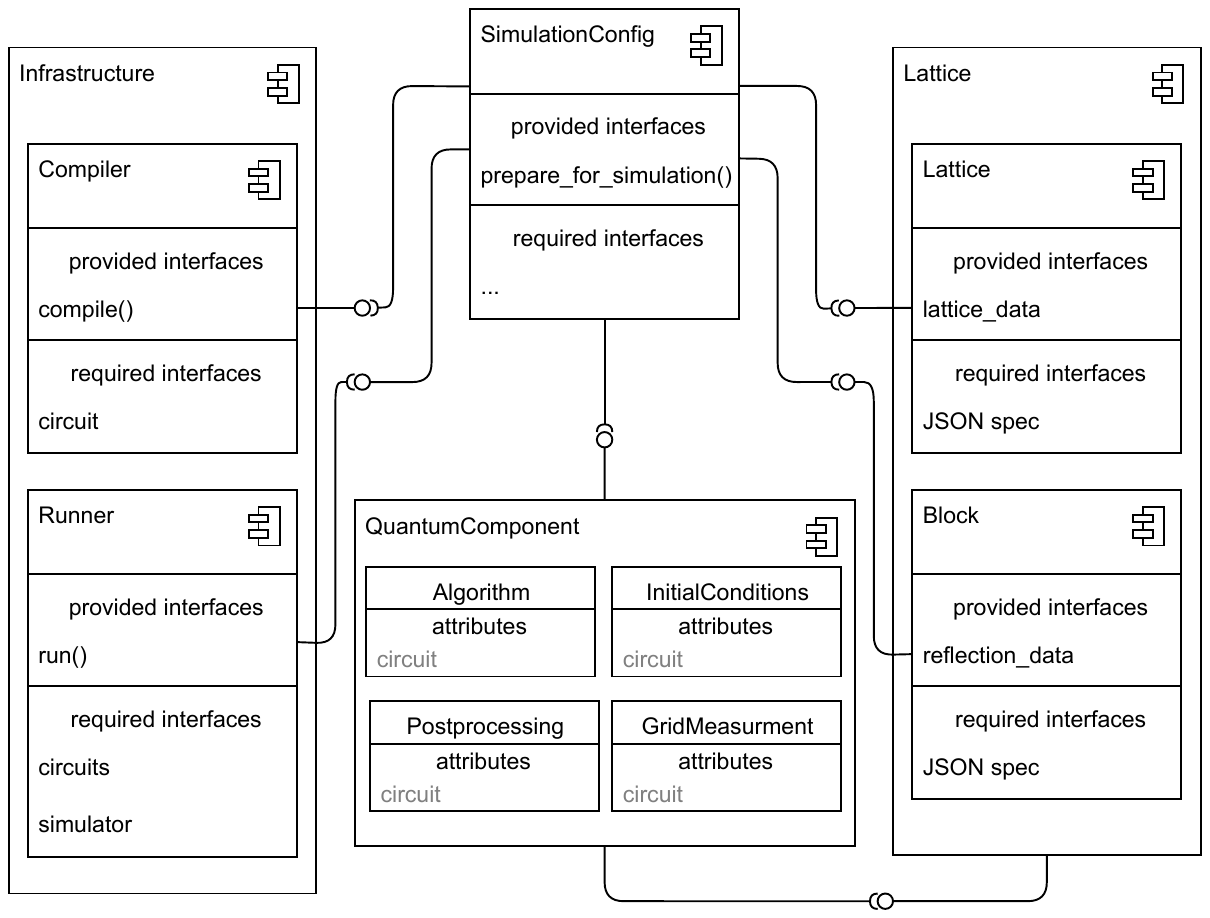}
	\vspace{-8cm}
	\caption{Representation of the system-wide \qlbm~architecture.}
	\label{fig:component-diagram}
\end{figure}

Integrating QBM circuits into broader
quantum software stacks is crucial for increasing the accessibility
of QBMs, as well as for expediting novel research in the NISQ era.
We design the architecture of the \qlbm~framework
around facilitating the use of the quantum components described
in the previous subsection.
\Cref{fig:component-diagram} gives an overview
of the three main components of \qlbm, as well as how
they come together to enable seamless user interaction.

At the bottom of the figure, the quantum components
of the QBM are mostly isolated from the remainder of the framework.
To further decouple the quantum circuit logic from the surrounding
utilities, we introduce a \texttt{Lattice}
component which handles specification parsing and preprocessing.
This module is depicted on the right hand-side of \Cref{fig:component-diagram}
and is tasked with the conversion of high-level specification
into information that can parameterize the construction of quantum circuits.
This includes parsing geometry specification
in the \texttt{Block}  class for the application
of boundary condition and determining the appropriate (minimal)
register setup for simulating the system in the \texttt{Lattice} class.
Quantum components use this information to construct appropriate circuits
on a per-algorithm basis.

External infrastructure is again handled in isolation
to encourage modularity and extendibility.
The \texttt{Infrastructure} component contains
both \texttt{Compiler} and \texttt{Runner}
classes that wrap the circuit transpilation utilities available
in Qiskit \cite{qiskit2024} and Tket \cite{sivarajah2020t}.
These utilities enable resource estimation experiments
for different hardware specifications, including
variable gate sets and qubit connectivities.
Uniform interfaces make it easy for the user to access
those services without requiring low-level tuning
of the underlying libraries.

Finally, the detached components are brought together
in an interface called a \texttt{SimulationConfig}.
This class ties together the algorithmic components
that make up a QBM, as well as additional
simulation options the user can configure.
Such items include the preferred compiler and simulator,
as well as their specific parameters.
The appeal of this highly coupled interface
is that it automates the process of preparing
the high-level quantum circuits generated in the component
module for execution on a specific quantum or classical hardware platform.
This allows the entire bundle of quantum circuits simulation parameters
to be forwarded to the \texttt{Runner} in one go.
\Cref{subsec:interfacing-integration} provides an
example of how the entire end-to-end simulation
workflow can be performed in just a few lines of code.

To complement modularity and isolation, the architecture of \qlbm~promotes
a high standard of code quality and reproducibility.
In addition to the open-source access, \qlbm~contains an extensive
suite of over 200 unit, integration, and end-to-end tests,
which target both low-level implementation details (such as geometry parsing),
as well as high-level features such as compatibility with several Qiskit simulators.
Supplementary to unit tests, \qlbm~hosts
a comprehensive documentation website with dozens of examples,
thousands of lines of in-code comments,
and additional tutorials that delve into advanced
applications of the software aimed at developing novel algorithms.

\subsection{Performance Enhancements \label{subsec:performance-enhancements}}

In an era without fault-tolerant quantum hardware, 
classical hardware plays a crucial role in accelerating
quantum algorithm research.
To do this effectively, we require
simulation software that enables classical hardware
to emulate quantum computers in the first place.
In this section, we highlight two directions that can enhance the performance
of QBM algorithms in an environment dominated by quantum simulation.
The first direction is algorithmic.
This includes any improvements that can be made
to quantum circuit design, as well as any computations
that can be delegated to classical information processing instead
of relying on the exponentially more expensive computation of a quantum state.
The second direction is computational.
This direction consists of exploits
that classical hardware allows, which would otherwise be
physically impossible on quantum computers.
Efficient statevector manipulations 
are an example of such an optimization.
In quantum computing, the no-cloning theorem 
prohibits exact copies of statevectors and measurement
often requires exponentially many shots of a circuit.
Such limitations can be circumvented in simulations.
Effective implementations of such techniques are
crucial for accelerating research into QBMs, as they save
researchers invaluable time and computational resources.
We first describe examples of algorithmic improvements in 
\Cref{subsec:algorithmic-imporvements} before addressing
their computational counterparts in \Cref{subsec:computational-imporvements}.

\subsubsection{Algorithmic Improvements \label{subsec:algorithmic-imporvements}}

Algorithmic improvements concern techniques that
reduce the complexity of quantum circuits in either
depth, number of qubits, or total number of gates.
Here, we refer again to the QTM
algorithm developed by \citet{schalkers2024efficient}
as an example, and highlight
two techniques that help reduce circuit complexity,
while focusing on how \qlbm~facilitates such improvements.

\paragraph{Ancilla qubit reuse.} Effectively leveraging the state
of ancilla qubits can help reduce both 
the number of gates and qubits
required to perform certain computations.
Here, we give two examples that curtail
gate and qubit requirements, respectively.
We first consider the depth of QTM algotrihm's streaming operator.
The algorithm leverages an ancilla system that determines whether populations of
particles stream in a given timestep subdivision, as computed by a CFL counter.
Discrete velocities that should stream in a substep are identified
in the quantum state through ancilla qubits $a_{v,i}$ that pertain to whether
particles with a specific discrete velocity $v$ are streamed in dimension $i$.
A naive implementation would first perform the streaming operation, reset the
state of the ancilla qubits, and then compute the boundary condition operator on
the resulting state.
However, since populations that have not streamed in the
CFL substep are not affected by boundary conditions, the same
ancilla state can be re-used to control which velocity
directional qubits are inverted by boundary conditions.
\Cref{fig:cqbm-code} exemplifies this optimization,
where the \texttt{StreamingAncillaPreparation} operator
is only used at the end of the CFL iteration, and not between 
each streaming and boundary condition routine.
The indexing methods of \qlbm's \texttt{Lattice}
classes allow for the seamless utilization of qubits
without manually performing the tedious indexing operations that change
with each system and lattice discretization.

Ancilla qubits can also be effectively reused to reduce the memory
requirements of heterogeneous boundary conditions.
Consider again \Cref{fig:cqbm-code} and the application of
both specular and bounceback boundary conditions.
Specular reflection entails the reversal of the velocity normal
to the reflection surface, whereas bounceback reflection
requires that all directions are inverted,
irrespective of the contact surface.
To practically realize specular reflection,
we use $d$ ancilla qubits to determine
which dimensions a population has reached an object in, 
which enables the computation of the velocity components to invert.
For bounceback reflection,
a single ancilla qubit suffices to determine whether particles
have exited the fluid domain, which triggers the reversal of all
directional velocity qubits \citep{schalkers2024momentum}.

A straightforward implementation of a system that supports
geometry with heterogeneous boundary conditions would therefore
utilize $d + 1$ ancilla qubits.
In the \qlbm~implementation of the QTM algorithm,
we only require the $d$ ancilla qubits that specular reflection necessitates.
This is made possible by imposing the restriction that the domains of
specular reflection and bounceback objects are separated
by at least two grid points.
If this constraint is satisfied, multiple obstacles with
either reflection method can leverage the same qubits without
causing any interference within the quantum state.
Though marginal for ideal fault-tolerant computers,
such improvements prove significant for classical emulation.
The \texttt{Lattice} class registers again make such optimizations
trivial to implement by allowing shared access
to all qubit registers from inside primitives and operators.
Moreover, the straightforward chaining of operators
makes it easy to compose circuits based on intermediate states,
and each \texttt{Lattice} class can incorporate
specification parsers that warn users if constraints are violated.

\paragraph{Adaptable register setup.} To further exploit the properties
of specific QBM setups, \qlbm~allows for a flexible register setup
that minimizes resource requirements.
The same observation that allows ancilla qubits to be reused
also leads to the simulation requiring fewer qubits for the simulation
of the end-to-end QTM algorithm.
As the \texttt{Lattice} object parses the input specification,
it tracks the different kinds of boundary conditions present in the system.
If the system only contains bounce-back boundaries,
the register is automatically shrunk to only require one 
ancilla qubit that suffices to perform reflection.
The relative indices of the remainder qubits are automatically
adjusted such that the users does not need to manually
adjust the circuits not affected by this change.
If, however, the system contains mixed boundary conditions, the
register is widened to accommodate the $d$ ancillae that
specular reflection utilizes, and the mechanism described
previously commences to effectively reuse the available qubits
for both boundary conditions.

\paragraph{Classical logic computation.} We again use the same
specular reflection operator of the QTM to highlight how the delegation
of logic to classical preprocessing can effectively
speed up simulation.
We give two examples of preprocessing techniques that simplify
both the quantum circuits and their simulation.

First, we consider specular reflection against a wall
in the nominal case, where particles do not encounter a corner of the object.
To reflect particles in a way that is physically correct, the
quantum circuits needs to determine which of the velocity directions to invert.
While this is simple to do in classical LBMs, the quantum circuit
additionally requires the information to persist within the state \emph{after}
particles have been removed from the non-fluid domain, as to reset the
state of any ancilla qubits that would otherwise
later interfere with the computation.
There are multiple ways to implement this logic.
Detecting and reseting such states
can be achieved both in the quantum circuit by means of extra ancilla qubits,
as well classically by manually defining which
velocity directions each wall surface affects.
However, the former requires significant quantum resources
and expensive additional logic, while the latter
is error-prone and difficult to debug.

In \qlbm, we provide an alternative implementation
that performs this computation in terms of automated
boolean logic operations in a step that precedes the assembly
of the quantum circuit.
Specifically, we take advantage of the encoding of velocities in the QTM algorithm.
We leverage this encoding by formulating a boolean
function over spacial properties of the object's edges,
that provides the information required to invert
and reset the appropriate velocity qubits.
To achieve this, we define near-corner points
as $2d$-dimensional boolean vectors in the cartesian product
space described by \Cref{eq:point-definition}.

\begin{equation}
    \mathbf{point} = \mathbf{bound} \times \mathbf{outside}
    \label{eq:point-definition}
\end{equation}

Here, $\mathbf{bound}, \mathbf{outside} \in \{\top, \bot \}^d$
are $d$-bit structures that encode the position of near-corner points per dimension.
The $\mathbf{bound}$ property encodes whether the point belongs
to a surface that is a lower ($\bot$) or an upper ($\top$) bound
of the object.
The $\mathbf{outside}$ values denote whether the point
is outside ($\top$) or inside ($\bot$) the object bounds.
Since both variables are dimension- and position-agnostic,
their cross product produces a data structure that encodes the
position of \emph{each} near-corner point of any cuboid-shaped object.
To determine whether an ancillary qubit is to be inverted
after performing the reflection step, \qlbm~simply queries
a single boolean value per dimension.
This value is computed as given in \Cref{eq:velocity-inversion}.

\begin{equation}
    \mathbf{inversion} = \mathbf{bound} \otimes \mathbf{outside}
    \label{eq:velocity-inversion}
\end{equation}

Where $\otimes$ corresponds to the point-wise \texttt{XOR} function.
That is, each ancillary qubit state is reset controlled
on (1) the position of the gridpoint within the lattice
and (2) the inversion boolean value associated with its relative
position with respect to the object.
This technique is powerful for two reasons.
First, it saves $d$ qubits that would be required
to implement a quantum counterpart to this computation without altering
the other components of the quantum state.
Second, since all classical computations required
for this purpose are trivial object-agnostic boolean operations,
the cost associated with this method is negligeable with
respect to the rest of the algorithm.
\qlbm~enables such computations to be carried out entirely independently
from the quantum circuit generation details, and implements them in a separate
\texttt{Block} class, which interfaces with the \texttt{Lattice} counterpart.
In practice, this means users can choose to tune the reliance of their methods
on classical computation without necessitating
any change to previously implemented circuits.
We note that \qlbm~uses the same mechanism to generate the reflection
circuits for both 2D and 3D reflection circuits, including
all edge cases around cuboid objects.
The cartesian product of \Cref{eq:point-definition} generalizes
to both points and edges (in 3D), and while the specific function
used to assign inversion boolean values differs per case,
its implementation remains straight-forward and efficient.

\subsubsection{Computational Improvements \label{subsec:computational-imporvements}}

Computational improvements involve techniques that
leverage current classical hardware to simulate
QBM circuits efficiently.
In this section, we outline three kinds of techniques
tailored to exploit the structure of QBM algorithms.

\paragraph{Statevector snapshots.}

Lattice Boltzmann Methods are inherently time-dependent algorithms.
In both classical and quantum LBMs, computations
occur in a temporal loop that consists of repeated steps.
This means that the circuits that implement
QBMs may be similar or even identical for each individual time step.
On real quantum hardware, this observation is of lesser importance.
While circuits can be reused, the statevector produced by
a QBM circuit after one step cannot be cloned
as input for the next.
However, quantum simulators available today do offer this option.
In this subsection, we show how taking advantage
of the availability of the entire statevector
can drastically decrease the time required to simulate QBMs.

\begin{figure}
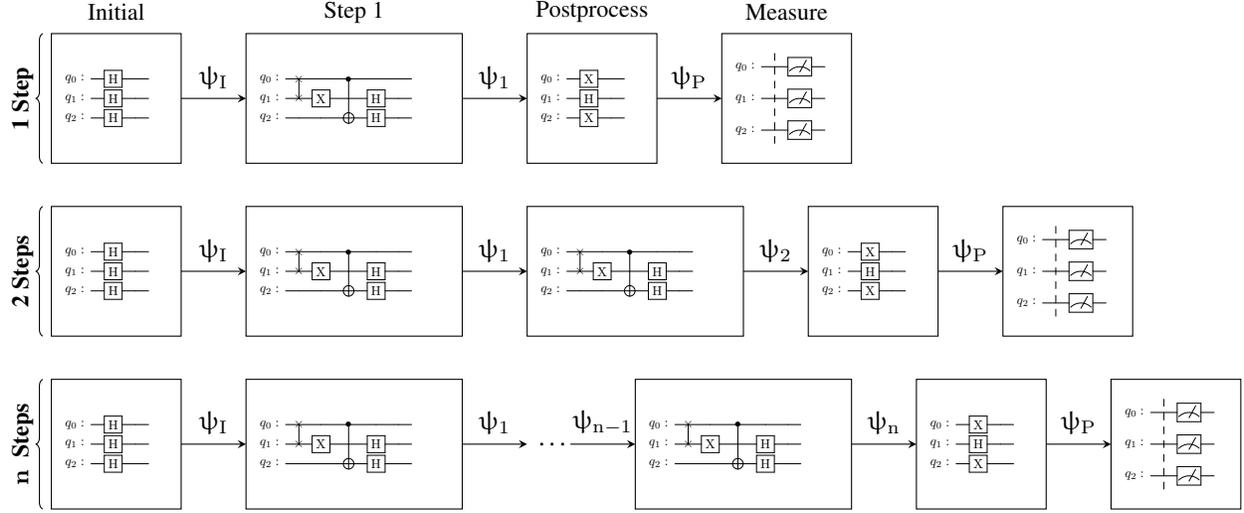
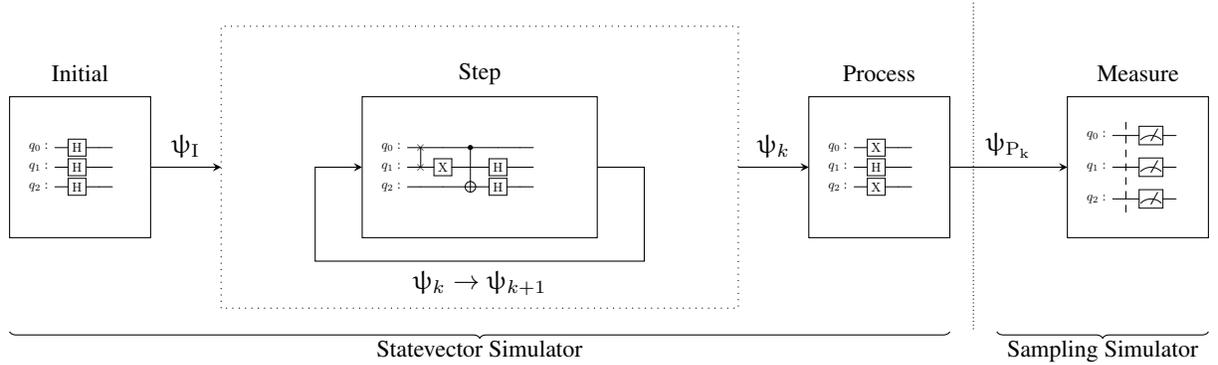

    \centering
    \hfill
    \subcaptionbox{Standard simulation of an $n$-step QBM algorithm.\label{fig:statevector-snapshots-off}}{\begin{tikzpicture}[scale=1.15]{node/.style={circle, draw, thick}}

\draw [decorate, decoration = {brace,mirror}]  (-0.85, 0.75) -- node[pos=0.5,above,rotate=90] {\footnotesize \textbf{1 Step}} (-0.85,-0.75);

\node[] at (0, 1)   (a) {\footnotesize Initial};
\draw[] (-0.75, -0.75) rectangle (0.75, 0.75);
\node[text width=2cm] at (0, 0) {
    \scalebox{0.5}{
        \input{diagrams/initial_conditions_circuit}
    }
};

\draw [-stealth] (0.75, 0) -- node[pos=0.5,above]{$\uppsi_\mathrm{I}$} (1.5, 0);

\node[] at (2.75, 1)   (a) {\footnotesize Step 1};
\draw[] (1.5, -0.75) rectangle (4, 0.75);
\node[text width=2cm] at (2.25, 0) {
    \scalebox{0.5}{
        \input{diagrams/simple_circuit}
    }
};

\draw [-stealth] (4, 0) -- node[pos=0.5,above]{$\uppsi_1$} (4.75, 0);

\node[] at (5.5, 1)   (a) {\footnotesize Postprocess};
\draw[] (4.75, -0.75) rectangle (6.25, 0.75);
\node[text width=2cm] at (5.5, 0) {
    \scalebox{0.5}{
        \input{diagrams/postprocessing_circuit}
    }
};

\draw [-stealth] (6.25, 0) -- node[pos=0.5,above]{$\uppsi_\mathrm{P}$} (7, 0);

\node[] at (7.75, 1)   (a) {\footnotesize Measure};
\draw[] (7, -0.75) rectangle (8.5, 0.75);
\node[text width=2cm] at (7.75, 0) {
    \scalebox{0.5}{
        \input{diagrams/measurement_circuit}
    }
};


\draw [decorate, decoration = {brace,mirror}]  (-0.85, -1.25) -- node[pos=0.5,above,rotate=90] {\footnotesize \textbf{2 Steps}} (-0.85,-2.75);
\draw[] (-0.75, -2.75) rectangle (0.75, -1.25);
\node[text width=2cm] at (0, -2) {
    \scalebox{0.5}{
        \input{diagrams/initial_conditions_circuit}
    }
};

\draw [-stealth] (0.75, -2) -- node[pos=0.5,above]{$\uppsi_\mathrm{I}$} (1.5, -2);

\draw[] (1.5, -2.75) rectangle (4, -1.25);
\node[text width=2cm] at (2.25, -2) {
    \scalebox{0.5}{
        \input{diagrams/simple_circuit}
    }
};

\draw [-stealth] (4, -2) -- node[pos=0.5,above]{$\uppsi_\mathrm{1}$} (4.75, -2);

\draw[] (4.75, -2.75) rectangle (7.25, -1.25);
\node[text width=2cm] at (5.5, -2) {
    \scalebox{0.5}{
        \input{diagrams/simple_circuit}
    }
};

\draw [-stealth] (7.25, -2) -- node[pos=0.5,above]{$\uppsi_\mathrm{2}$} (8, -2);

\draw[] (8, -2.75) rectangle (9.5, -1.25);
\node[text width=2cm] at (8.75, -2) {
    \scalebox{0.5}{
        \input{diagrams/postprocessing_circuit}
    }
};

\draw [-stealth] (9.5, -2) -- node[pos=0.5,above]{$\uppsi_\mathrm{P}$} (10.25, -2);

\draw[] (10.25, -2.75) rectangle (11.75, -1.25);
\node[text width=2cm] at (11, -2) {
    \scalebox{0.5}{
        \input{diagrams/measurement_circuit}
    }
};


\draw [decorate, decoration = {brace,mirror}]  (-0.85, -3.25) -- node[pos=0.5,above,rotate=90] {\footnotesize $\mathbf{n}$ \textbf{ Steps}} (-0.85,-4.75);
\draw[] (-0.75, -4.75) rectangle (0.75, -3.25);
\node[text width=2cm] at (0, -4) {
    \scalebox{0.5}{
        \input{diagrams/initial_conditions_circuit}
    }
};

\draw [-stealth] (0.75, -4) -- node[pos=0.5,above]{$\uppsi_\mathrm{I}$} (1.5, -4);

\draw[] (1.5, -4.75) rectangle (4, -3.25);
\node[text width=2cm] at (2.25, -4) {
    \scalebox{0.5}{
        \input{diagrams/simple_circuit}
    }
};

\draw [-stealth] (4, -4) -- node[pos=0.5,above]{$\uppsi_\mathrm{1}$} (4.75, -4);

\node[text width=2cm] at (5.75, -4) {\ldots};

\draw [-stealth] (5.25, -4) -- node[pos=0.5,above]{$\uppsi_\mathrm{n-1}$} (6, -4);

\draw[] (6, -4.75) rectangle (8.5, -3.25);
\node[text width=2cm] at (6.75, -4) {
    \scalebox{0.5}{
        \input{diagrams/simple_circuit}
    }
};

\draw [-stealth] (8.5, -4) -- node[pos=0.5,above]{$\uppsi_\mathrm{n}$} (9.25, -4);

\draw[] (9.25, -4.75) rectangle (10.75, -3.25);
\node[text width=2cm] at (10, -4) {
    \scalebox{0.5}{
        \input{diagrams/postprocessing_circuit}
    }
};

\draw [-stealth] (10.75, -4) -- node[pos=0.5,above]{$\uppsi_\mathrm{P}$} (11.5, -4);

\draw[] (11.5, -4.75) rectangle (13, -3.25);
\node[text width=2cm] at (12.25, -4) {
    \scalebox{0.5}{
        \input{diagrams/measurement_circuit}
    }
};

\end{tikzpicture}}%
    \hfill%
    \subcaptionbox{Snapshot simulation of an $n$-step QBM algorithm.\label{fig:statevector-snapshots-on}}{\begin{tikzpicture}[scale=1.25]{node/.style={circle, draw, thick}}

\node[] at (0, 1)   (a) {\footnotesize Initial};
\draw[] (-0.75, -0.75) rectangle (0.75, 0.75);
\node[text width=2cm] at (0, 0) {
    \scalebox{0.5}{
        \input{diagrams/initial_conditions_circuit}
    }
};

\draw [-stealth] (0.75, 0) -- node[pos=0.5,above]{$\uppsi_\mathrm{I}$} (1.5, 0);

\draw[style=dotted] (1.5, -1.5) rectangle (7, 1.5);
\node[] at (4.25, 1)   (a) {\footnotesize Step};
\draw[] (3, -0.75) rectangle (5.5, 0.75);
\node[text width=2cm] at (3.75, 0) {
    \scalebox{0.5}{
        \input{diagrams/simple_circuit}
    }
};

\draw [-stealth] (5.5, 0) -- (6, 0) -- (6, -1) -- node[pos=0.5,below]{$\uppsi_k \to \uppsi_{k+1}$} (2.5, -1) -- (2.5, 0) -- (3, 0);

\draw [-stealth] (7, 0) -- node[pos=0.5,above]{$\uppsi_{k}$} (7.75, 0);

\node[] at (8.5, 1)   (a) {\footnotesize Process};
\draw[] (7.75, -0.75) rectangle (9.25, 0.75);
\node[text width=2cm] at (8.5, 0) {
    \scalebox{0.5}{
        \input{diagrams/postprocessing_circuit}
    }
};

\draw [-stealth] (9.25, 0) -- node[pos=0.5,above]{$\uppsi_\mathrm{P_k}$} (10.5, 0);

\node[] at (11.25, 1)   (a) {\footnotesize Measure};
\draw[] (10.5, -0.75) rectangle (12, 0.75);
\node[text width=2cm] at (11.25, 0) {
    \scalebox{0.5}{
        \input{diagrams/measurement_circuit}
    }
};

\draw[densely dotted] (9.5, 1.75) -- (9.5,-1.75);
\draw [decorate, decoration = {brace,mirror}]  (-0.75, -1.75) -- node[pos=0.5,below] {\footnotesize Statevector Simulator} (9.25,-1.75);

\draw [decorate, decoration = {brace,mirror}]  (9.75, -1.75) -- node[pos=0.5,below] {\footnotesize Sampling Simulator} (12,-1.75);

\end{tikzpicture}}%
    \caption{Comparison of simulation strategies for multistep QBM algorithms.}
    \label{fig:statevector_snapshots}
\end{figure}

To showcase this improvement, we consider a scenario
in which the goal of the simulation is to perform
$n$ time steps of the QTM algorithm,
and visualize the entire flow field encoded
in the quantum state \emph{after each step}.
This is a common method that
helps researchers verify whether the implementation of the circuit
produces physically consistent behavior.
\Cref{fig:statevector-snapshots-off} depicts what an
implementation of this workflow might look like on quantum hardware.
Each time step requires a different quantum circuit,
that is made up of $k$ repetitions of the same
circuit nestled between state preparation and post-processing primitives.
Following each execution, measurements collapse the quantum state
onto basis states that can reconstruct the flow field.
In total, the QTM single-time step quantum circuit is executed
$\mathcal{O}(n^2)$ times, not accounting for the multiple shots required
for each step.
This scaling emerges as a consequence of the fact
that to simulate and approximate the
flowfield over $n$ time steps, all $1 \leq k \leq n$
time steps require a separate simulation of $k$ concatenated
single-step circuits each.
Therefore, the single-step circuit is executed $n(n+1)/2$ times,
not accounting for the number of shots at each step.
This is a general requirement of the task, rather than a consequence
of the specific algorithmic implementation.

Fortunately, quantum simulators afford the extraction of additional
information from their representation of quantum states without
requiring multiple shots.
\Cref{fig:statevector-snapshots-on} depicts an efficient implementation
of the same workflow on quantum simulators.
Though the same structure is preserved, the transfer
of information from one time step to another is fundamentally different
in \qlbm's implementation.
While iterating through the (identical) time-step circuits,
each statevector $\uppsi_k$ produced by the circuit
undergoes the following process.
If post-processing is required, \qlbm~first creates a copy of the statevector,
which it then passes on to a sampling subroutine.
Otherwise, a single instance of the data suffices.
Afterwards, \qlbm~produces samples of
this statevector \emph{without} changing it.
This enables the extraction of information required for visualization
while maintaining the statevector intact.
After the information has been extracted, the same
statevector is fed back into the same time step
circuit, which updates it with one more iteration.

The core difference between the two approaches is that the latter
effectively takes \emph{snapshots} of the statevector as 
the $n$ timestep circuits iteratively evolve it.
This enables a drastic increase in performance,
as only $\mathcal{O}(n)$ simulations of the single step
algorithm are required.
In \qlbm, we additionally increase the efficiency
of this method by only constructing and transpiling
the time step circuit once and re-using it
for each iteration.
In practice, using this method to simulate novel
developments of circuits massively decreases
the time researchers spend verifying their implementations.

\paragraph{Statevector sampling.}

Quantum simulation is an area of active
research that continuously improves the performance of
quantum emulation through new methods and software.
Emerging methods all have different strengths and weaknesses,
which depend on the underlying hardware and the simulated quantum circuits.
To take advantage of the plethora of simulation paradigms, we split the simulation procedure into two distinct phases, as illustrated in \Cref{fig:statevector-snapshots-on}.

The first phase consists of the simulation of the
quantum circuit up to but not inlcuding the post-processing step.
To enable the snapshot-driven execution that reduces the complexity order,
the simulator that performs this routine must be able to
retrieve the entire statevector at the end of each time step.
The second phase of the procedure involves postrocessing and measurement.
Unlike the time step circuit(s), post-processing and measurement circuits
are generally shallow and simple.
In addition to this, the state that emerges as a result
of the post-processing circuit is not of any importance to the
next iteration of the algorithm, and instead
serves visualization and verification purposes.
Because the two phases are constrained
by significantly different requirements,
\qlbm~allows for the specification of different simulators for each phase.

The goal of this distinction is to take advantage of simulation technology
that favors the simulation of shallow circuits
in the latter stages of each time step.
In the general case, this requires two copies of the statevector
to be kept in memory at the same time -- one that
serves as input to the following time step iteration, and one
that serves as input to the post-processing and measurement stage.
Fortunately, in the nominal scenario where researchers are
interested in visualizing the entire flow field evolved by the circuits,
post-processing is circumvented.
In practice, this means that the quantum state computed by the time step circuit
can effectively be \emph{sampled} by a different, more suitable simulator
with no additional copy required.

\paragraph{Reinitialization.}

Simulating QBM algorithms on classical hardware comes
with inherent limitations.
Clearly, the exponential memory advantage that quantum computing
promises is not achievable through classical emulation.
With this fundamental limitation come several practical challenges.
One such challenge facing the snapshot and sampling techniques stems
from the transition between time steps.
For algorithms like QTM, the time step transition
on quantum simulator consists of a single straightforward
transfer of the statevector from one circuit to the next.
However, this is not the case for most QBM algorithms.

We consider the STQBM algorithm
\cite{schalkers2024importance} as an example.
To circumvent the non-unitarity of streaming in the computational
basis state encoding, the STQBM algorithm uses velocity information
from neighboring gridpoints.
In the general case, this utilizes $\mathcal{O}(N_t^d)$
additional qubits to propagate spacial information in time,
with $N_t$ the number of time steps to simulate and
$d$ the number of dimensions of the problem.
Due to memory limitations, the direct simulation of more than a few
time steps of this algorithm is infeasible for most classical hardware available today.
To circumvent this constraint, an effective \emph{reinitializaiton}
mechanism is required.
Such a mechanism converts information encoded in the quantum state
at the end of a simulation into a quantum circuit
that prepares the state for following time step(s).

We emphasize that this use of reinitialization is not
exclusive to the STQMB.
Though the underlying reasons differ, other approaches may require similar
mechanisms.
We consider the LCU-based paradigm 
\citep{budinski2021quantum, ljubomir2022quantum} as an additional example.
Due to how LCU-based methods perform collision, the quantum state contains
an additional component, which is orthogonal to the component
encoding the state of the flow field.
In practice, this makes many measurements obtained
from the quantum state produced by LCU time step circuits
irrelevant for the flow field computation.
This makes reinitializaiton techniques valuable,
as they segregate the algorithm into
smaller restart-driven blocks that prevent
the orthogonal state component from propagating.

\begin{figure}
	\centering
	\input{diagrams/reinitialization}
	\caption{\qlbm~reinitialization loop.}
	\label{fig:reinitialization}
\end{figure}

To facilitate the development of all three kinds of QBMs,
we equip \qlbm~with a uniform reinitailization interface.
\Cref{fig:reinitialization} depicts how reinitialization 
integrates into the efficient QBM simulation loop.
After a circuit implementing one or more time steps has been assembled
(upper, left-hand quadrant), simulation can commence.
A simulator backend of the user's choice, such as Qiskit or Qulacs
then evolves the quantum state by the full time step
circuit from $\ket{0}^{\otimes n}$ into $\uppsi_k$ (upper right-hand quadrant).
The nominal \qlbm~flow  then directs the quantum state
towards the sampling backend.
At this stage, information is extracted from the state in the form
of \emph{counts}, which include the measured basis states and their relative frequency
with respect to a pre-determined number of shots.
It is this information that enables \qlbm's integration
with external visualization techniques.

After extracting the samples from the quantum state,
the statevector and the counts are fed to an instance of the
\texttt{Reinitializer} class.
This class provides restart methods that are tailored to the algorithm being simulated.
The \texttt{Reinitializer} object performs the appropriate processing
of the input data to obtain the initial conditions of the next time step circuit.
For the QTM algorithm, this only involves wrapping the statevector
in an appropriate interface -- counts are disregarded.
For STQBM, the process involves parsing the counts information and constructing
a new quantum circuit that propagates the velocity information
to neighboring grid points.
In this case, the statevector object is discarded.
Following the reinitialization step, the novel initial conditions
circuit is automatically compiled to the appropriate target platform.
To enable seamless interaction between the reinitializer and the rest
of the simulation infrastructure, \qlbm's base \texttt{Reinitializer}
provides a stable and uniform interface that requires no modification
for the implementation of novel algorithms.
Together with the flexible component interface described in \Cref{subsec:internal-architecture},
this enables the on-the-fly composition of the newly derived circuit
with the unchanged time step circuit from the previous iteration.
We note that next to developing the quantum circuits,
implementing (or reusing) a \texttt{Reinitializer} class
and adjacent \texttt{Result} class (for parsing and visualizing samples)
are the only other step researchers need to take to fully implement a QBM
algorithm in \qlbm, while retaining all performance enhancements.

\subsection{Interfacing and Integration \label{subsec:interfacing-integration}}

Providing intuitive interfaces is crucial for
making software accessible to both researchers and practitioners.
In this section, we describe how users can interact with \qlbm~as
a QBM simulation tool, before addressing \qlbm's integration
with external software.

\subsubsection{Interfacing}

One of the advantages of \qlbm's internal component representation
is that it enables automatic circuit construction.
This shifts the burden of realizing system-specific circuits
from the user to the logic inside the software.
In turn, this means users should have access to a seamless way
of specifying complex quantum circuits.
To address this need, \qlbm~provides the option for users to specify
system properties in an implementation-agnostic
way through a \texttt{JSON} interface.

\Cref{fig:qlbm-json-spec} contains an example of such a specification.
When parsing this specification, \qlbm~uses the lattice
properties to determine the appropriate qubit register setup,
as well as the structure, position, and order of quantum components
that compose the algorithm.
Next to discretization details
such as the number of gridpoints in each dimension (lines 3-6)
and number of discrete velocities (lines 7-10),
users can additionally specify properties
of the geometry within the system (lines 12-23).
Each geometric object is composed of a lower and an upper bound
in each dimension (lines 14, 15), together with the object's boundary conditions (line 16).
In the current version of \qlbm, only cuboid-shaped geometries
are supported for the QTM algorithm, with either specular or bounce-back boundary conditions.
To accomodate the development of novel algorithms, \qlbm~parses the exact same
specification file to derive multiple QBMs.
Internally, the \texttt{Lattice} class provides a base that includes
a parser, which specialized implementations can leverage.
In practice, this means utilities that warn users of ill-formed specifications
can be shared between algorithms, and excessive information 
(i.e., boundary conditions that are not yet suported by some QBMs) can be discarded.

\begin{figure}
	\centering
	\input{code/specification}
	\caption{Sample \qlbm~lattice configuration.}
	\label{fig:qlbm-json-spec}
\end{figure}

\qlbm~offers several alternatives that bridge the gap between
the high-level \texttt{JSON} specification and nuanced
details of the simulation.
\Cref{fig:qlbm-simulation-code} depicts the most user-friendly 
interface available in \qlbm, based around the \texttt{SimulationConfig} wrapper.
First, users choose a lattice file to simulate,
written in the same format as \Cref{fig:qlbm-json-spec} (line 12).
Next, the \texttt{SimulationConfig} class (lines 13-23) defines
a convenient container that bundles together all required simulation data.
This includes the specific components that make up the quantum algorithm (lines 14-17),
the platforms that run and compile the quantum circuits (lines 18-20),
and explicit simulator choices (lines 21-23).
The components and simulator choices correspond exactly to the workflow
described in \Cref{subsec:computational-imporvements} and \Cref{fig:statevector_snapshots}.
At this stage, no additional user configuration is required,
as \qlbm~infers all quantum registers and circuits based on the parsed lattice data alone.

Following configuration, a single call to the \texttt{prepare\_for\_simulation()}
method of the configuration object determines whether the user-supplied
configuration is valid and compiles all circuits to the appropriate simulator format.
Next, users need only make a call to the
\texttt{run()} method of a \texttt{Runner} object,
specifying the number of time steps to simulate,
the number of shots to sample from the
statevector, and whether to use the snapshot mechanism.
We note that the distinction between where the sampling and snapshot mechanisms
are specified stems from the fact that sampling
requires a different compilation pipeline if enabled,
and as such needs to be specified at circuit assembly time.
We discuss the different available options for both compilers and runners
in the following subsection.

\begin{figure}
	\centering
	\input{code/runtime}
	\caption{Sample \qlbm~usage.}
	\label{fig:qlbm-simulation-code}
\end{figure}

\subsubsection{Infrastructure and Integration \label{subsec:infrastructure-integration}}

The field of quantum software is rapidly evolving.
The quality, scope, and variety of available
software are continuously increasing as researchers develop
new methods to bridge the gap between the current-day hardware
and fault tolerance.
Such improvements are evident at multiple
stages of the quantum software pipeline,
and taking advantage of them is crucial 
for increasing the pace and quality of related research.
In this section, we elaborate how advances in quantum software technology
affect \qlbm, and how its integration with external software infrastructure
can accelerate QCFD research.
We begin by addressing how \qlbm~assembles quantum circuits
before focusing on simulation, compilation, and visualization, respectively.

\paragraph{Circuit specification.}
Over the years, many quantum programming frameworks and languages
have emerged for various platforms and specifications.
Popular general-purpose quantum programming frameworks include the
Open Quantum Assembly Languange (OpenQASM) \cite{cross2022openqasm}
Quipper \citep{green2013quipper},
ProjectQ \citep{steiger2018projectq},
Cirq \cite{cirq2024}, and
Qiskit \citep{qiskit2024}.
\qlbm~builds its internal representation of quantum circuits
on top of Qiskit's \texttt{QuantumCircuit} class.
Specifically, each \qlbm~primitive, operator, and algorithm
holds an internal Qiskit quantum circuit that is built from
either a small set of parameters or from a \texttt{Lattice} specification.
We select Qiskit for three main reasons.
First, its large ecosystem encompasses many
useful pieces of adjacent infrastructure,
including analysis and visualization tools.
Second, Qiskit's popularity
increase the accessibility and reach of \qlbm~with a broader user base.
Finally, its rich toolkit of circuits makes the specification
of elaborate quantum circuits seamless.
It is especially this feature that enables
\qlbm's modular component architecture
and circuit composition capabilities.

\paragraph{Simulation.}

Using classical hardware to simulate the exponentially
large space that logical qubits reside in is an inherently limiting task.
In the case of arbitrary random circuits, the amount of classical memory
required to simulate a quantum algorithm doubles with every qubit.
In spite of this fundamental limitation, researchers and engineers have been
developing tools that can significantly accelerate quantum simulation
and increase the domain of algorithms that classical hardware can meaningfully emulate.
When considering the advancements that quantum simulation technology has 
undergone in recent years, one can distinguish between two main directions.
The first direction concerns the simulation method.
Quantum states can be represented directly
as numerical instances of statevectors and density matrices,
but also symbolically through graphs \citep{viamontes2005graph},
tensor networks \cite{orus2019tensor},
and decision diagrams \cite{wecker2014liqui}.
Though none of these methods fully overcome the exponential
disadvantage that classical hardware faces, each of them
may provide practically meaningful advantages for
particular classes of problems.
Second, improvements in simulation performance can also come from
how the simulation method integrates with hardware.
To this end, engineers have developed simulators
for different platforms,
including general purpose CPUs, 
as well as ARM-based clusters \cite{imamura2022mpiqulacs}
and GPUs \cite{gutierrez2007cuda, efthymiou2021qibo}.

Selecting suitable simulation methods
for the hardware at hand is pivotal for speeding up the simulation
and development of QBM algorithms.
To reduce the friction between the zoo of available simulator
implementations and the researchers looking to exploit them,
\qlbm~provides several built-in presets.
The simulation modules of \qlbm~rely on two external libraries:
Qiskit \citep{qiskit2024} and Qulacs \cite{suzuki2021qulacs}.
Qiskit provides a plethora of simulator options 
through its \texttt{qiskit-aer} package, which include
nine different simulation techniques for heterogeneous
computer hardware.
Qulacs provides a competing implementation of a
statevector simulator that has been shown to
outperform Qiskit in several benchmarks \citep{suzuki2021qulacs}.
Both Qiskit and Qulacs provide CPU and GPU implementations of their
methods under similar interfaces, which \qlbm~inherits and provides
different installations for.
A third option natively supported in \qlbm~is MPIQulacs \citep{imamura2022mpiqulacs},
a multi-node alternative of Qulacs designed for ARM-based
compute clusters, that enables the distributed simulation of quantum algorithms.

The implementations that link \qlbm~to each simulator reside under
specialized implementations of a base \texttt{Runner} class.
Any of the three implementations can be swapped into the same
workflow that \Cref{fig:qlbm-simulation-code} shows 
(\ie by swapping the \texttt{QiskitRunner} instruction in line 27).
Moreover, users can decide which of the three (CPU, GPU, or MPIQulacs)
options they want to use at install time by specifying a single
installation parameter.
To further boost the reproducibility of research carried out with \qlbm,
we provide versioned installation options for python environments,
as well as Docker containers.
For the latter, we bundle \qlbm~with custom container images
that build on top of lightweight Python images 
and the NVIDIA cuQuantum Appliance for GPU simulation.

\paragraph{Circuit compilation.}

Much like classical software, high-level descriptions of quantum
circuits undergo a compilation process that translates the circuits
to an instruction set that is compatible with
a specific piece of hardware or simulator.
To complement the modularity of \qlbm's circuit specification
and simulation options, a similarly versatile compilation system is required.
Compilers with configurable platforms are called \emph{retargettable}.
We integrate \qlbm~with two such retargettable compilers:
Qiskit \citep{qiskit2024} and Tket \citep{sivarajah2020t}.
Both frameworks offer a broad range of compilation targets
and optimization techniques that can target both quantum simulators,
as well as quantum hardware.
Providing both options by default inside \qlbm~allows
researchers to experiment with competing compilation options
that may be favorable for different scenarios.
For this purpose, we additionally build logging
and analysis tools within \qlbm, that allow users
to automate the exploration and benchmarking of available options.

From an implementation standpoint, \qlbm~simplifies the
interaction between the framework and available compilers by providing
a single entry point through a \texttt{CircuitCompiler} class.
Through a single call to the object constructor, users can
specify the compiler platform (Qiskit or Tket)
and its target (Qiskit or Qulacs) without any additional complications
(\ie \texttt{CircuitCompiler("TKET", "QISKIT")}).
Additional options are available through the single \texttt{.compile()} method
that allows users to select different backends and optimization levels.
Currently, \qlbm~makes all cross-combinations of compiler platforms
and targets available.
This includes all extensions of Qiskit and Tket \texttt{Backend}s,
both simulator and hardware-specific emulators,
as well as the Qulacs and MPIQulacs.
To further simplify the interaction between the \qlbm~users
and the vast number of viable combinations, we implement compilers
in such a way that \qlbm~components can be directly
input into the compiler, without requiring any processing 
or decomposition of the circuit
(\ie \texttt{CircuitCompiler("QISKIT", "QULACS").compile(CQLBM(lattice), ...)}).

\paragraph{Visualization.}

Visualization serves two main purposes in \qlbm.
The first is to convert the information extracted
from the quantum state at the end of the computation
into a visual interpretation of the flow field.
Developers can use this feature to verify
the correctness of an implementation and the evolution
of the flow field over time.
For debugging purposes, we additionally allow users
to save the statevector and counts to disk
alongside the flow field visualization.
The second goal of \qlbm~visualization tools
is to provide a means for quickly assessing
the difference in the performance and scaling
of QBM algorithms and their adjacent infrastructure.
To achieve this, \qlbm~provides scripts
that automate both
the parsing of log information into common data formats and
the conversion of the extracted data into plots.

Implementation-wise, flow field visualization
relies on the Visual Toolkit (VTK) \citep{schroeder2000visualizing}
software package to efficiently encode flow field count data into
standard formats.
For geometry data, \qlbm~converts the cuboid bounds
into triangulated surfaces in the commonplace \texttt{stl} format.
Both flow field and geometry visualization conversions take place
into specialized implementations of the base \texttt{QLBMResult} class
that is specific to each QBM and integrates with the rest of
the infrastracture.
We select these formats specifically such that
each artifact generated by \qlbm~can be visualized
in Paraview \cite{paraview} without any additional user intervention.
Finally, performance logs are parsed by scripts bundled in Jupyter Notebooks
for easy editing and plotting within the \qlbm~environment.
\section{Results \label{sec:results}}

This section demonstrates the experimental capabilities of \qlbm.
We highlight the computational and analysis tools of \qlbm~in
order of the workflow depicted in \Cref{fig:qlbm-workflow}.
\Cref{subsec:algo-res} showcases how the properties
of quantum circuits can be analyzed by parameterizing
the high-level \texttt{JSON} configuration files.
In \Cref{subsec:compiler-res}, we address
the next step in the \qlbm~workflow -- compilation.
We specifically analyze the performance of different compilers
and the trade-offs they present.
\Cref{subsec:sim-res} covers simulation performance in detail.
We begin by comparing the performance of different simulation platforms,
before considering GPU compatibility.
We also highlight how the computational improvements of
concerning statevector processing significantly speed up simulation.
Finally, \Cref{subsec:visual-res} displays
the visualization options that \qlbm~supports.
All experiments were performed on a machine equipped with an AMD
Ryzen 7 5800H CPU, 16 GB of RAM, and an NVIDIA 3050Ti GPU with 4GB of VRAM.

\subsection{Algorithmic scalability \label{subsec:algo-res}}

We begin analyzing the scalability of the 
QTM algorithm \citep{schalkers2024efficient}
expressed as the high-level circuit that \qlbm~generates
as a platform-agnostic quantum circuit.
The simple specifications through which users configure
\qlbm~makes it such that many different parameters
can be isolated and analyzed independently.
For the purposes of this experiment, we select the number of objects
within the fluid domain and the number of grid points in each dimension
as the subjects of the analysis,
because of their impact on the structure and depth of the quantum circuit.
We consider the depth of the circuit and the time it takes \qlbm~to assemble it.
To mitigate the effects of noise, we execute each experiment 5 times.

We note that the QTM algorithm requires $\sum_j \lceil \log_2 N_{g_j} \rceil$
positional qubits, $\sum_j \lceil \log_2 N_{v_j} \rceil$ velocity qubits, and
$4d-2$ ancilla qubits, where
$d$ is the number of spatial dimensions of the system,
$N_{g_j}$ is the number of gridpoints the lattice spans across dimension $j$,
and $N_{v_j}$ is the number of discrete velocities in that dimension.
Throughout our experiments, we select algorithms with either $2$ or $3$
dimensions and between $16\times16$ and $512\times512\times512$ gridpoints,
which require between $18$ and $43$ qubits,
which may be reduced by either $1$ or $2$ by the adaptive register setup mechanism.
For an in-depth analysis of the qubit register setup
and of the complexity of the algorithm,
we point the reader to Sections 3 and 7 of \cite{schalkers2024efficient}, respectively.

\begin{figure}
    \centering
    \hfill
    \subcaptionbox{Duration of circuit assembly of 2D QTM algorithm.\label{fig:res-1-algo-duration-2d}}{\includegraphics[scale=0.43]{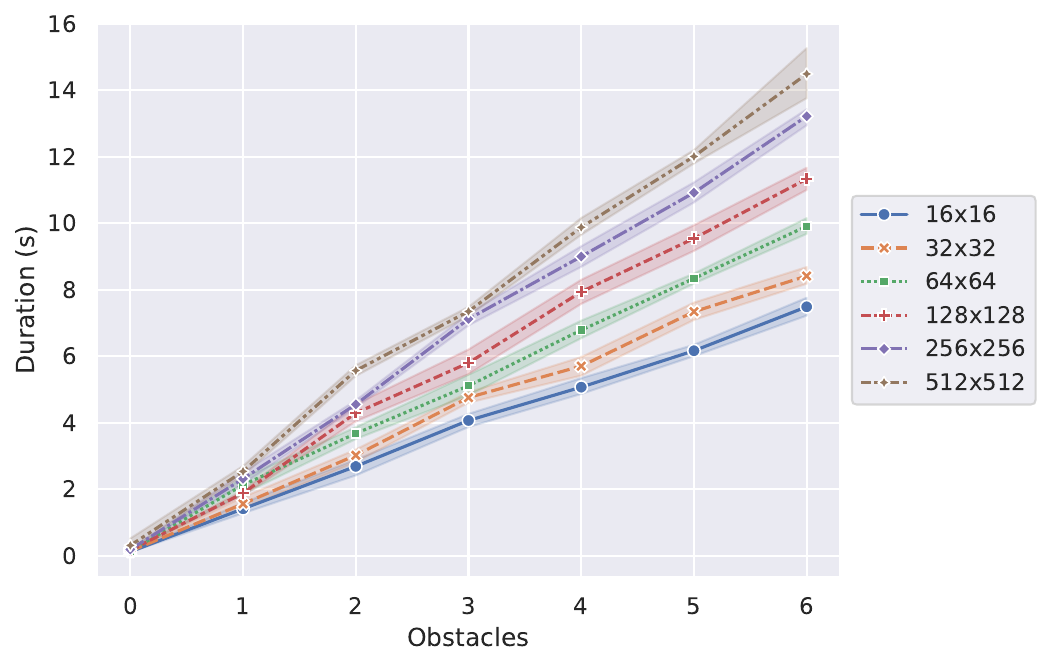}}%
    \hfill%
    \subcaptionbox{Duration of circuit assembly of 3D QTM algorithm.\label{fig:res-1-algo-duration-3d}}{\includegraphics[scale=0.43]{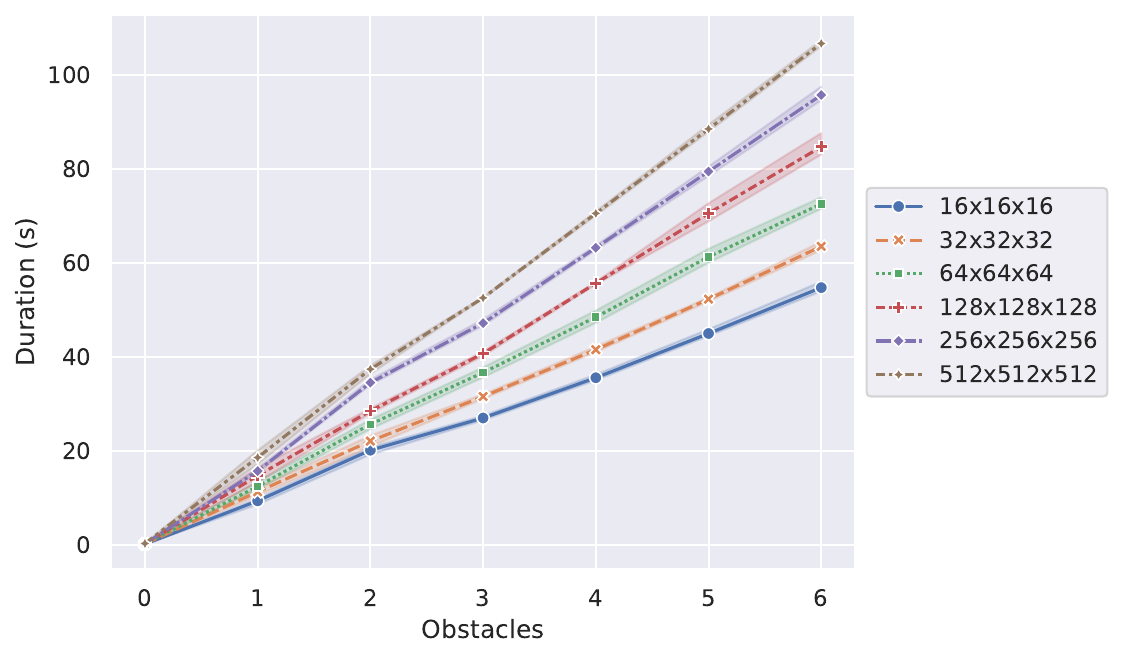}}%
    \\
    \hfill
    \subcaptionbox{Number of gates of 2D QTM algorithm.\label{fig:res-1-algo-gates-2d}}{\includegraphics[scale=0.43]{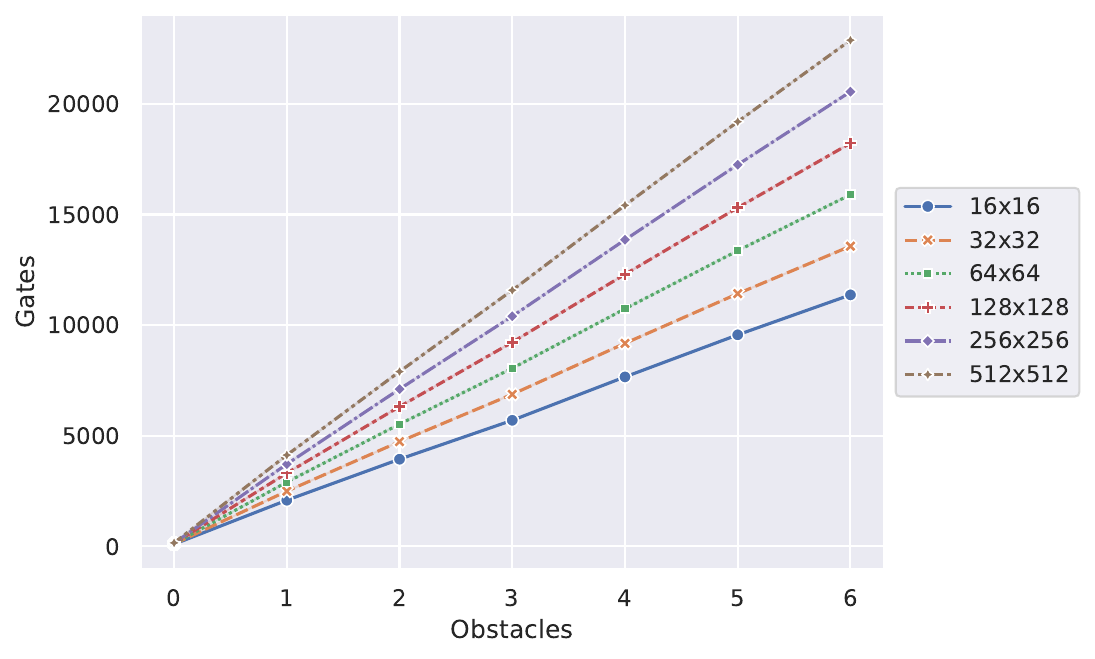}}%
    \hfill%
    \subcaptionbox{Number of gates of 3D QTM algorithm.\label{fig:res-1-algo-gates-3d}}{\includegraphics[scale=0.43]{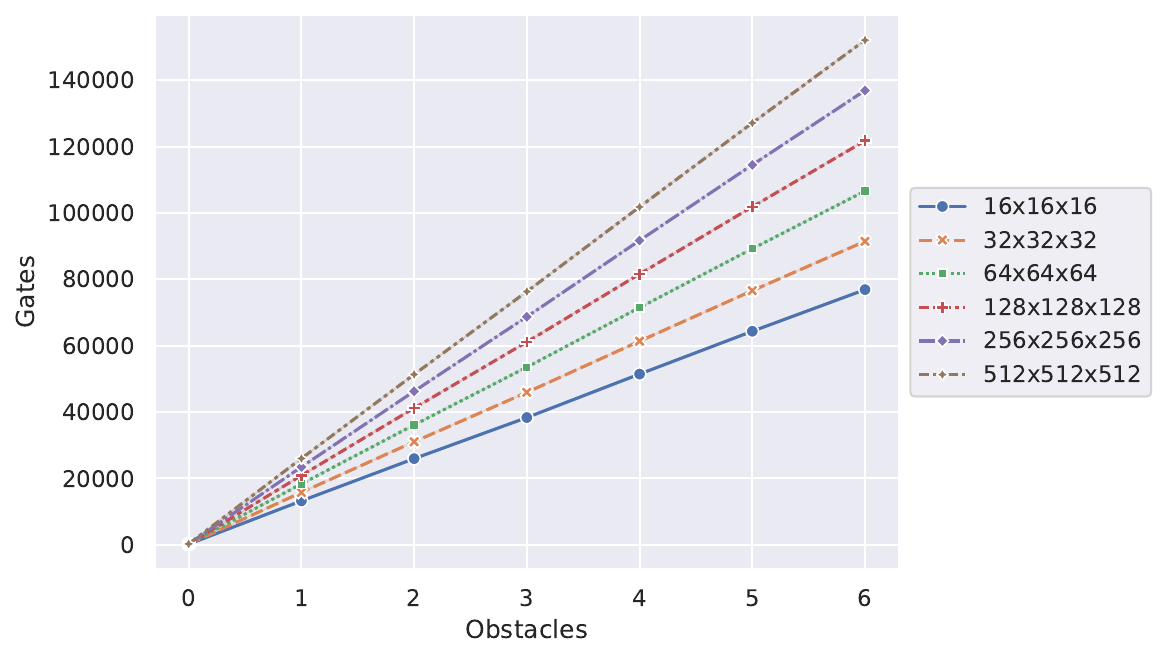}}%
    \caption{Comparison of QTM algorithm properties for uniform lattices between $16\times 16$ and $512\times 512 \times 512$ grid points and between 0 and 6 obstacles with bounce-back boundary conditions.}
    \label{fig:res-1-algo-scalability}
\end{figure}

\Cref{fig:res-1-algo-duration-2d} and \Cref{fig:res-1-algo-duration-3d}
show the how dimensionality, grid refinement, and geometry affect the
time it takes for \qlbm~to assemble the quantum circuits.
Dimensionality and grid refinement both affect the number of qubits
required to simulate the system, and as such they introduce
additional linear complexity in the circuit.
The number of (cuboid) obstacles has a similarly significant
and linear on the assembly duration.
This behavior is expected, as the implementation of QTM algorithm
reuses structurally identical comparator operators to determine which populations to stream,
for each edge and surface of an object.
Of note is that the majority of the complexity of the QTM algorithm stems from reflection.
Increasing the number of obstacles from 1 (median assembly time $5.35 \si{\second}$)
to 5 (median assembly time $34.8\si{\second}$) causes a similar
proportional increase in assembly time ($6.50$) as increasing
the dimensionality of the system from 2D (median assembly time $5.38$)
to 3D (median assembly time $37.4$) while keeping 
the number of obstacles fixed (proportional increase $6.96$).
The assembly duration increases are consistent with the number of gates
of the algorithm shown in \Cref{fig:res-1-algo-gates-2d} and \Cref{fig:res-1-algo-gates-3d}.
Both geometry complexity and grid refinement affect the scaling of the number of gates linearly.

The increase in both gate count and assembly time originates from two sources.
First, operations on individual lattice locations require
controlled operations based on the entire grid register.
These operations primarily occur around the corner points
of objects, and they make the distinction between which
populations are subject to boundary condition treatment.
As the size of grid register scales with the refinement
of the underpinning lattice, so does the number of gates
required to set and reset the quantum state for each point.
The second reason for the grid point-driven scaling has to do with the
controlled incrementation operation that both streaming and reflection utilize.
These circuits rely on a QFT operation followed by a controlled phase shift
that increments the position of particles in physical space by one grid point.
Each of these operations too scale with the size of the grid register,
as incrementation has to take place uniformly.
Geometry-based scaling stems from the fact that the \qlbm~implementation
of QTM boundary conditions iterates through each surface of each
obstacle in the lattice, which adds a number of gates that
scales linearly with the number of obstacles.
Finally, we emphasize that \qlbm~enables the analysis of such algorithmic properties
for all quantum components (\ie primitives, operators, algorithms), which in turn
facilitates resource estimation for different implementations.
In \Cref{subsec:compiler-res}, we extend this analysis to
low-level circuits targeted towards specific gate sets.

\subsection{Compiler comparison \label{subsec:compiler-res}}

We shift focus towards analyzing the QBM circuits
after transpiling them to lower-level gate sets.
Here, we consider the performance of the Qiskit and Tket compilers and their trade-offs.
We again use the end-to-end QTM algorithm \citep{schalkers2024efficient} as a benchmark, and
analyze three metrics for each compiler -- compilation time, circuit depth, and gate count.
Together, these three metrics give an indication
of the trade-offs that users face when choosing between
transpilation times and performance.
For the compiler platform, we select the Qulacs gate set available both in Qiskit
and Tket through the \texttt{qiskit-qulacs} and \texttt{pytket-qulacs} packages, respectively.
Qulacs has a significantly more restricted gate set than the one
that \qlbm~uses to construct quantum circuits,
which makes it a suitable candidate for such a benchmark
because of its likeness to real quantum hardware constraints.
We select a 17 qubit quantum circuit simulating one time step
of a $16 \times 16$ grid  with $4$ discrete velocities in each dimensions
and between $0$ and $6$ obstacles with bounce-back boundary conditions
placed at different positions on the grid.
We specifically select the number of obstacles
as the parameter to be varied as it only influences the depth and number of gates
of the circuit, rather than the number of qubits.
This factor also has the largest impact on algorithm complexity, after dimensionality.

\begin{figure}
    \centering
    \hfill
    \subcaptionbox{Circuit depth comparison.\label{fig:res-2-depth}}{\includegraphics[scale=0.33]{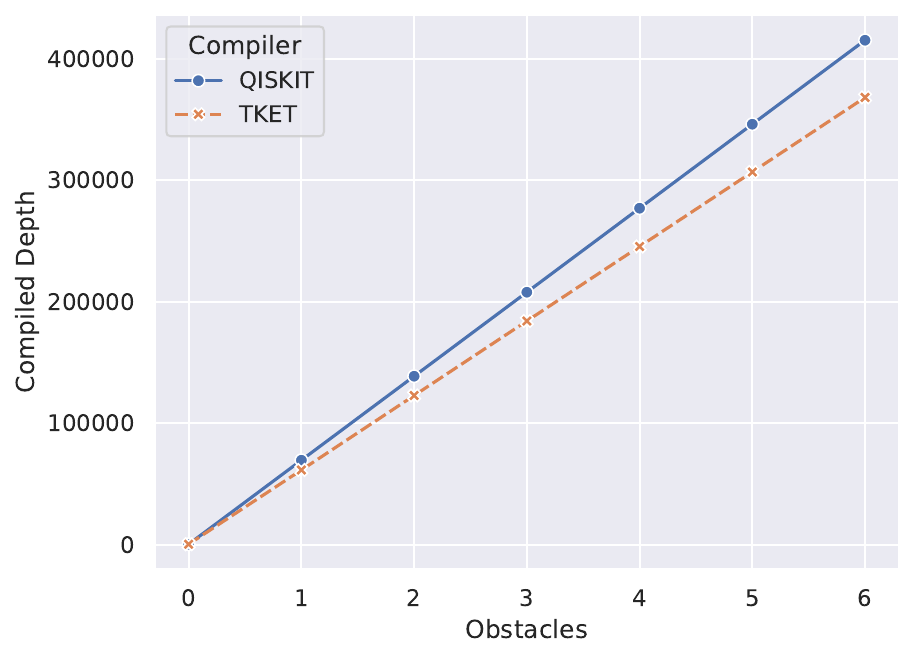}}%
    \hfill%
    \subcaptionbox{Gate count comparison.\label{fig:res-2-gates}}{\includegraphics[scale=0.33]{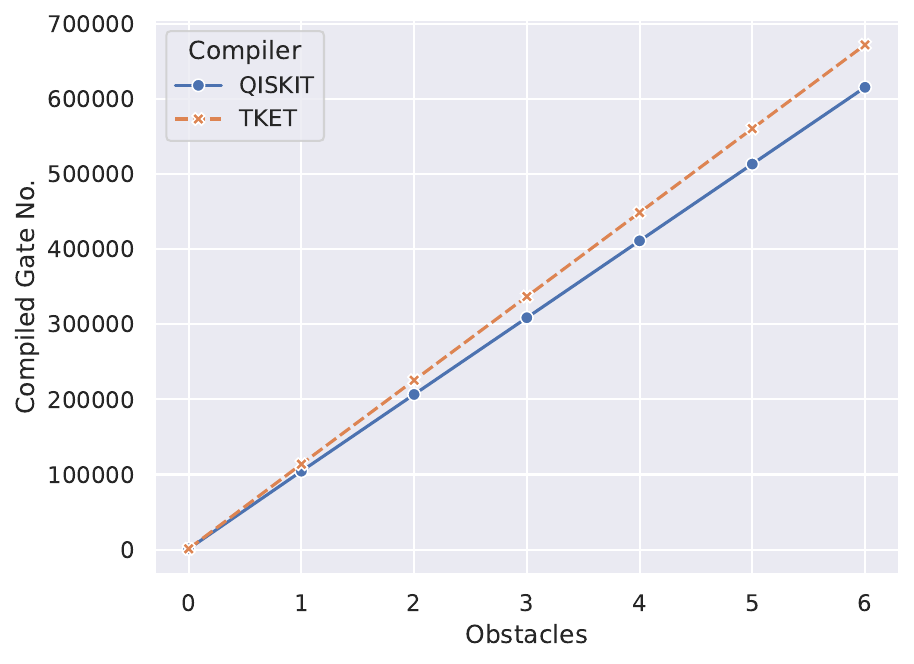}}%
    \hfill%
    \subcaptionbox{Compile time comparison.\label{fig:res-2-time}}{\includegraphics[scale=0.33]{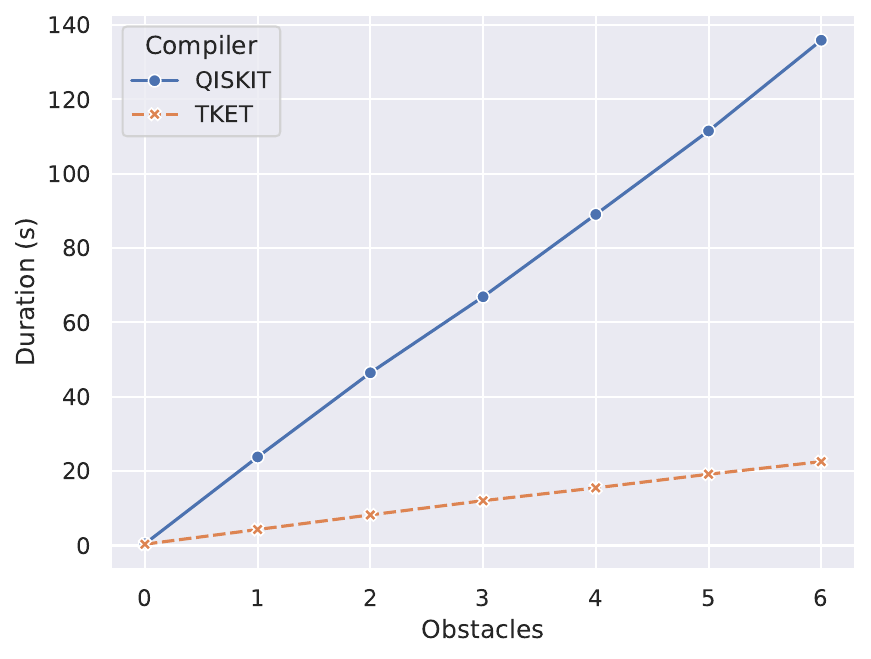}}%
    \caption{Compiler comparison for 2D QTM algorithm for a $16 \times 16$
    grid with 4 discrete velocities per dimension and between $0$ and $6$
    obstacles in the domain.}
    \label{fig:res-2-compiler-scalability}
\end{figure}

\Cref{fig:res-2-compiler-scalability} displays the results.
When assessing compiler performance in terms of the conciseness of the generated circuit,
\Cref{fig:res-2-depth} and \Cref{fig:res-2-gates} show that the Tket
and Qiskit compilers have different strenghts.
While Qiskit generates circuits that contain up to $50.000$ fewer gates
than their Tket counterparts, the Tket circuits
have a depth that is up to $20.000$ gates shallower.
For both compilers, the depth and the gate count scale linearly
with the number of obstacles in the grid, which is in line
with the scalability analysis.
A third natural consideration when assessing compilers is computation time.
While the scaling is again linear for both candidates,
Tket is significantly faster than Qiskit,
and is able to transpile the most complex circuit
in under one sixth of the time that Qiskit requires.
The \qlbm~analysis and benchmark suite makes 
such experiments easy to execute and replicate, which in turn helps
practitioners make informed decisions that can significantly accelerate their workflows.
We next extend this analysis to the performance of different 
simulators under nominal use cases on different hardware platforms.

\subsection{Performance comparison \label{subsec:sim-res}}

Selecting the appropriate simulation technology for the hardware
available at hand is a necessity for optimizing
the development process of novel QBMs.
Performance is sensitive to many factors,
including the simulation paradigm, its compatibility with available hardware,
and its suitability for the structure of the circuits being simulated.
Constructing QBM implementations that are versatile
enough to allow for experimentation with all of these parameters
is a time-consuming ordeal that \qlbm~seeks to relieve researchers of.
In this subsection, we demonstrate automated experiments
that users of \qlbm~can easily carry out to assess
the performance of simulation software for their specific needs.
We begin with assessing different simulators with out-of-the-box
settings before assessing the statevector snapshot
technique described in \Cref{subsec:computational-imporvements}
and showing its integration with GPUs.

\paragraph{Simulator comparison.}
Perhaps the most important choice when it comes to assessing simulation performance
is choosing the appropriate software library.
This poses a challenge for developers, as differences in library APIs,
software dependencies, and circuit assumptions can all hamper the simulation of QBMs.
To alleviate these burdens, \qlbm~provides two features.
First, the modular design of the \texttt{Runner} module allows
for easy extendibility to novel simulators.
Second, the built-in \texttt{SimulationConfig} class
automatically parses all components that make up the QBM into the appropriate
format for simulation, for any provided simulator.
We demonstrate such experiments by comparing the baseline Qiskit
\texttt{AerSimulator} with two alternatives: Qulacs \cite{suzuki2021qulacs} and DDSIM.
\footnote{DDSIM is available at \url{https://github.com/cda-tum/mqt-ddsim}.}
To ensure fairness, we compare each simulator against the baseline
in independent Python virtual environments because of dependency discrepancies.
Statevector snapshots and sampling are both turned off in this experiment.

\begin{figure}
    \centering
    \hfill
    \subcaptionbox{Comparison of Qiskit and Qulacs simulation performance.\label{fig:res-3-qulacs}}{\includegraphics[scale=0.43]{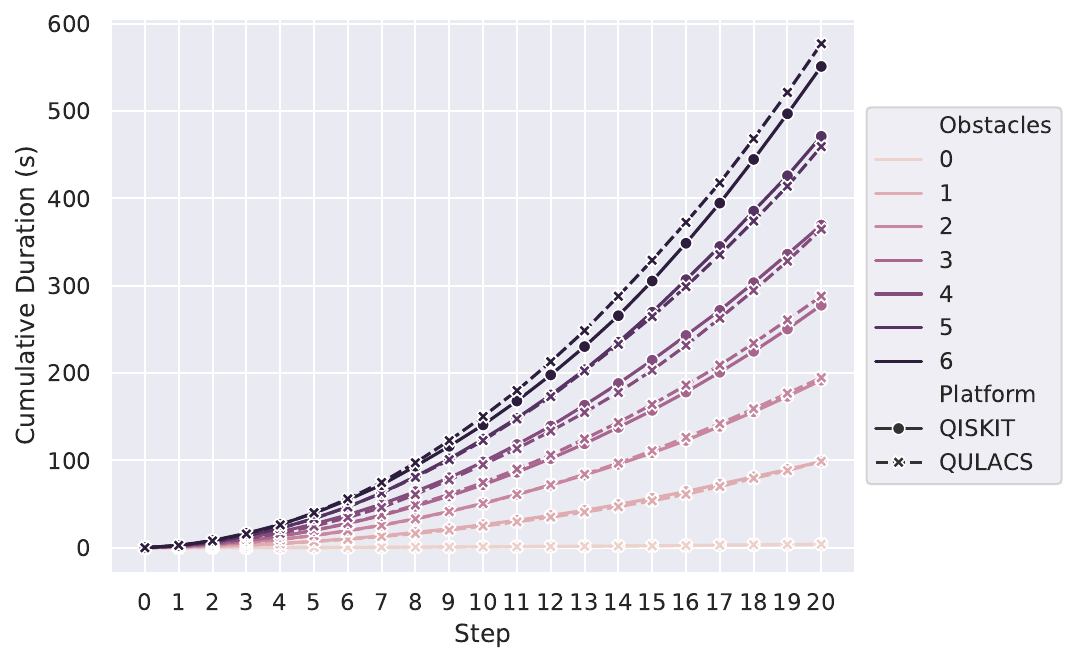}}%
    \hfill%
    \subcaptionbox{Comparison of Qiskit and DDSIM simulation performance.\label{fig:res-3-ddsim}}{\includegraphics[scale=0.43]{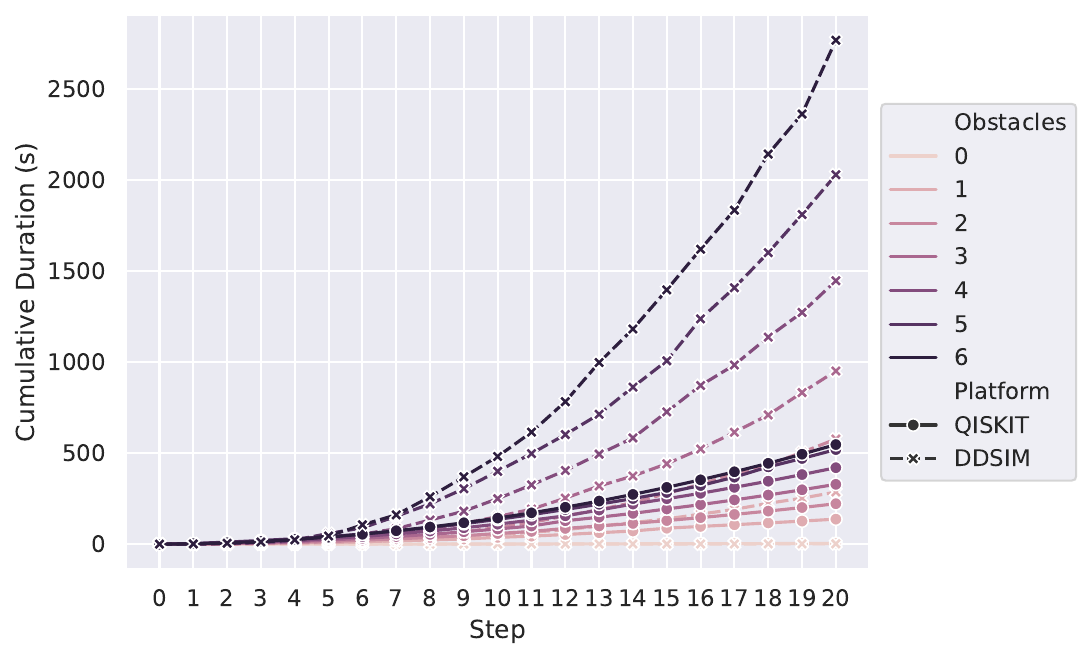}}%
    \caption{Simulator comparison for 2D QTM algorithm for a $16 \times 16$
    grid with 4 discrete velocities per dimension and between $0$ and $6$
    obstacles in the domain, for up to 20 time steps.}
    \label{fig:res-3-simulator-comp}
\end{figure}

\Cref{fig:res-3-simulator-comp} displays the results.
\Cref{fig:res-3-qulacs} indicates that Qiskit and Qulacs
perform similarly well for all $7$ instances.
For lattices with up to 3 obstacles, the performance
of the two simulators is almost indistinguishable.
For the instance with $4$ obstacles, Qulacs slightly outperforms Qiskit,
while the instances $3$, $5$, and $6$ obstacles slightly favor Qiskit.
\Cref{fig:res-3-ddsim} shows the comparison between the same Qiskit
simulator and DDSIM.
Here, all instances show a significant difference between the simulators,
in favor of Qiskit.
As the circuits grow more complex (\ie more obstacles), the difference
becomes practically more relevant.
While this in no way implies the general superiority of the Qiskit simulator,
it hints at the fact that the circuits that implement the QTM
algorithm in \qlbm~may be structurally a poor 
fit for the decision diagram decomposition that DDSIM relies on.
This kind of analysis can point researchers towards the simulator
that best fits the kind of algorithm they are working on extending or implementing.
In what follows, we analyze how the statevector snapshot technique
can signficiantly increase the performance of any simulator capable of
capturing entire statevectors.

\begin{figure}
    \centering
    \hfill
    \subcaptionbox{Comparison of statevector snapshot performance for up to 20 time steps of the QTM algorithm.\label{fig:res-3-statevector-snapshots-large}}{\includegraphics[scale=0.48]{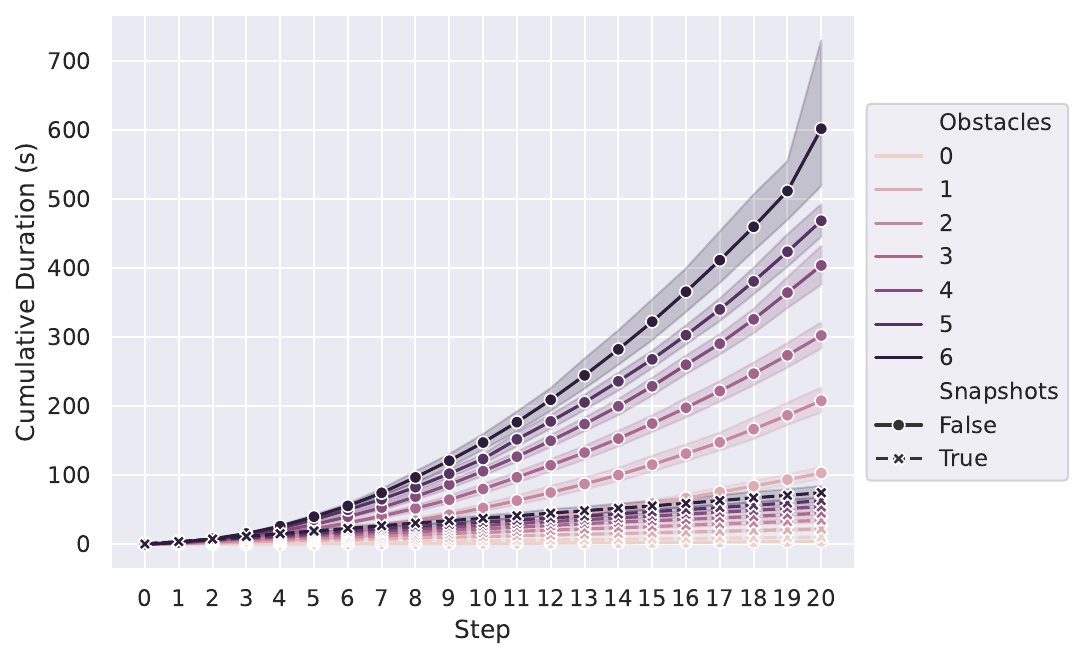}}%
    \hfill%
    \subcaptionbox{Comparison of statevector snapshot performance for up to 5 time steps of the QTM algorithm.\label{fig:res-3-statevector-snapshots-zoom}}{\includegraphics[scale=0.48]{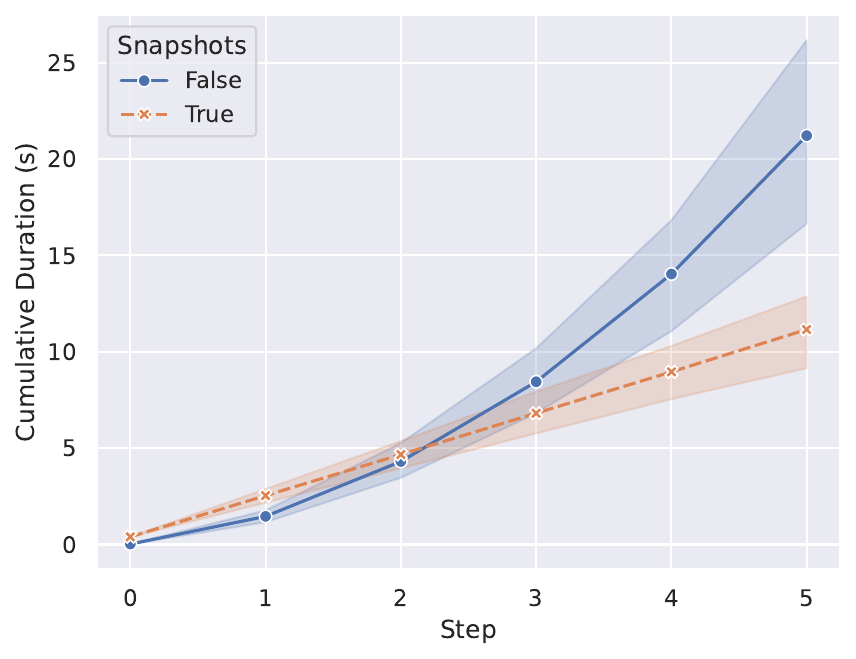}}%
    \caption{Comparison of statevector snapshot performance for a $16 \times 16$
    grid with 4 discrete velocities per dimension and between $0$ and $6$
    obstacles in the domain, for up to 20 time steps.}
    \label{fig:res-3-statevector-snapshots}
\end{figure}

\paragraph{Statevector snapshots.}
\Cref{fig:res-3-statevector-snapshots} shows the scalability of the
statevector snapshot and sampling techniques
described in \Cref{subsec:computational-imporvements}.
Dashed lines indicate configurations that were simulated
with both techniques enabled, whereas solid lines
indicate regular simulations.
Both sets of experiments were repeated 5 times, and the figures
illustrate the mean and standard deviation of the simulation time, respectively.
For consistency, we consider the same benchmark example as in the previous experiments.
Both the generation of the circuits, as well as the compilation (through Qiskit)
were handled through the standard \qlbm~workflow.
All simulations were carried out on Qiskit's
\texttt{AerSimulator} with the \texttt{statevector} method,
which has shown the best performance in previous instances.

The results confirm the complexity analysis provided in \Cref{subsec:computational-imporvements}.
Focusing on \Cref{fig:res-3-statevector-snapshots-large},
the results show how when combined, the snapshot and sampling techniques
can decrease the time required to perform a $20$-step simulation
by up to a factor of $6$.
The deviation in performance is increasingly visible as the complexity
of the circuit scales with the number of obstacles in the fluid domain.
Concretely, simulations that use both of our computational improvement techniques 
scale linearly in the number of steps simulated, while the standard
simulation method scales quadratically.
In the practical development cycle, this drastically
accelerates the pace at which researchers can verify
and debug their implementations.
Since the difference between statevector snapshots and regular simulations
scales linearly with the complexity of a single time step circuit,
the number of time steps that snapshots save is always higher for more complex systems.
This is a valuable improvement in practice, as more complex systems
generally require more runs to verify and debug.

We also note that the transfer of statevectors between simulators, despite
not requiring a deep copy, still introduces overhead that is meaningful in some instances.
This is especially visible for shorter circuits, where statevector transfer
between simulators takes up a higher percentage of the computational time.
For simpler circuits, such as the instances with $1$ and $2$ objects,
the scaling advantage only overtakes this overhead after $4$ and $3$ steps, respectively.
As the complexity of the simulated circuits increases however, the number
of steps required for the statevector snapshots to become advantageous decreases.
For the instance with $3$ obstacles, the overhead is only
unfavorable for the first time step, following which the scaling factor becomes dominant.
To better highlight this downside of snapshots, we zoom in in \Cref{fig:res-3-statevector-snapshots-zoom}.
Averaging over all lattice configurations, it takes $3$ time steps
to gain a practical advantage from the snapshot mechanism,
which also proved advantageous for all runs after $5$ time steps (or more).
As researchers are typically interested in simulating tens or hundreds of time steps
for algorithmic verification purposes, this overhead is rarely a downside in practice.
In the next paragraphs, we show the versatility of the snapshot mechanism by
performing simulations on a GPU.

\paragraph{GPU integration.}
All \qlbm~performance improvements can leverage
multiple compute architectures, including GPUs and ARM-based CPUs.
We demonstrate this by showing the applicability of the statevector
snapshot mechanism when applied to GPU simulation.
We use the same benchmark as in the previous two example
and compare the Qiskit \texttt{AerSimulator} running CPU and GPU devices
with both snapshots and sampling optimizations enabled.
The GPU simulator leverages the cuQuantum SDK \cite{bayraktar2023cuquantum}
and runs in a modified Docker container, based on the NVIDIA cuQuantum Appliance.
As with all examples used throughout this manuscript, we make
the container used in this benchmark available with the rest of code base.

\begin{figure}
    \centering
    \includegraphics[width=0.5\linewidth]{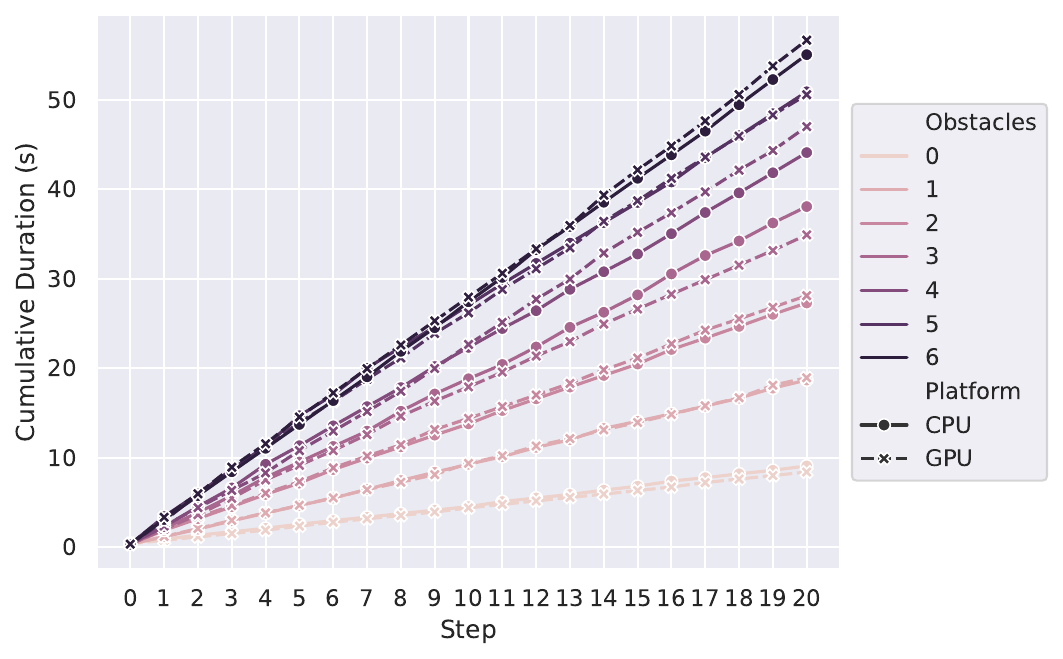}
    \caption{CPU and GPU performance comparison using statevector snapshots
    for a $16 \times 16$ grid with 4 discrete
    velocities per dimension and between $0$ and $6$
    obstacles in the domain, for up to 20 time steps.}
    \label{fig:res-3-gpu}
\end{figure}

\Cref{fig:res-3-gpu} shows the results.
Both the CPU and GPU simulator display the same
linear scaling as \Cref{fig:res-3-statevector-snapshots}.
The CPU version slightly edges its GPU counterpart in $4$ of $7$ instances,
but no significant difference occurs between the two.
As in the Qulacs and DDSIM example, these experiments
are meant to highlight the ease with which practitioners
can test different simulator options that pertain to
heterogeneous hardware platforms, within the \qlbm~workflow.
Furthermore, the results demonstrate the versatility
of the snapshot and sampling techniques, which nets users
significant improvements when compared to naive implementations.
In what follows, we feature how \qlbm~makes use of these effective
simulation techniques to create detailed and useful visualizations
of the system under simulation, concluding the workflow. 

\subsection{Visualization integration \label{subsec:visual-res}}

Visualization serves two important purposes in the \qlbm~pipeline.
First, it allows researchers to verify the correctness of their implementation.
This is especially important when addressing end-to-end algorithms.
Complete circuits may be hundreds of thousands of gates deep
and simultaneously address dozens of boundary condition edge cases,
and visualization provides a means of assuring that end-to-end integration
of the quantum components is sound in relation to the physical system being simulated.
The second grounds for visualization is accessibility.
For users familiar with classical CFD workflows,
integration with established visualization software
bridges the gap between the novelty of the quantum methods
and standard practices.
In this subsection, we demonstrate the built-in \textsc{Paraview} \cite{paraview}
integration of \qlbm~for both the QTM \cite{schalkers2024efficient} and
STQBM \cite{schalkers2024importance} algorithms.

\begin{figure}
    \centering
    \hfill
    \subcaptionbox{The initial conditions of the particle distribution. \label{fig:res-4-eight-00}}{\includegraphics[scale=0.22]{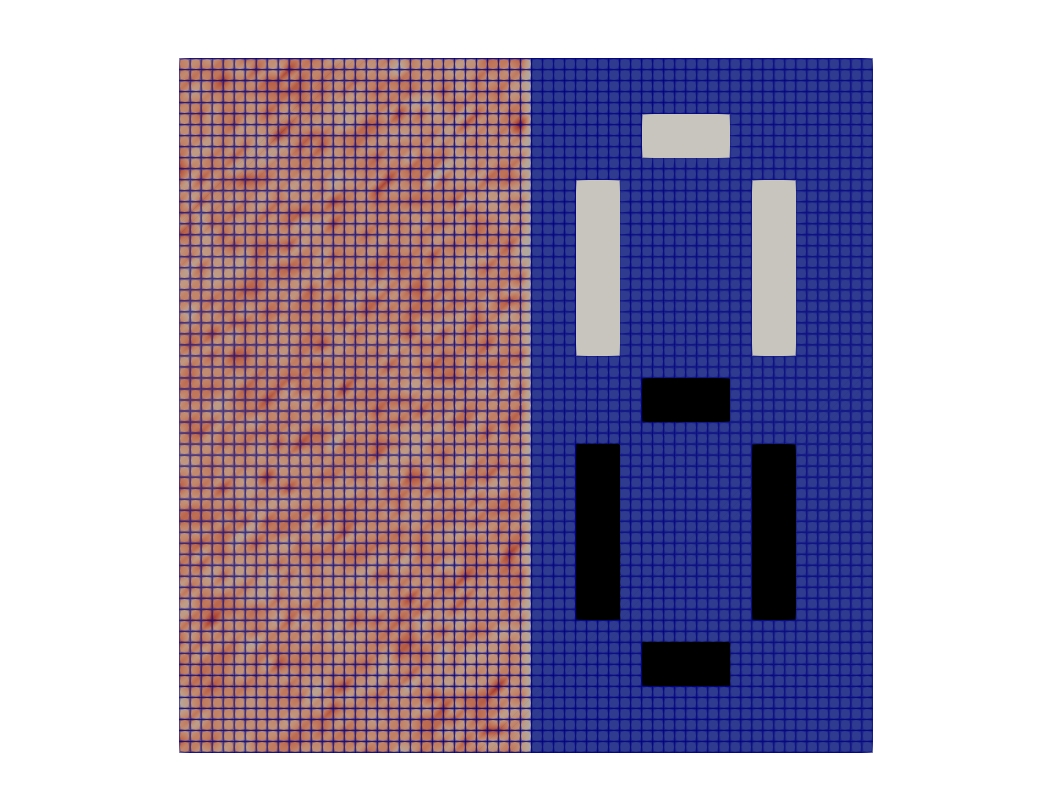}}%
    \hfill%
    \subcaptionbox{The system after 16 time steps.\label{fig:res-4-eight-16}}{\includegraphics[scale=0.22]{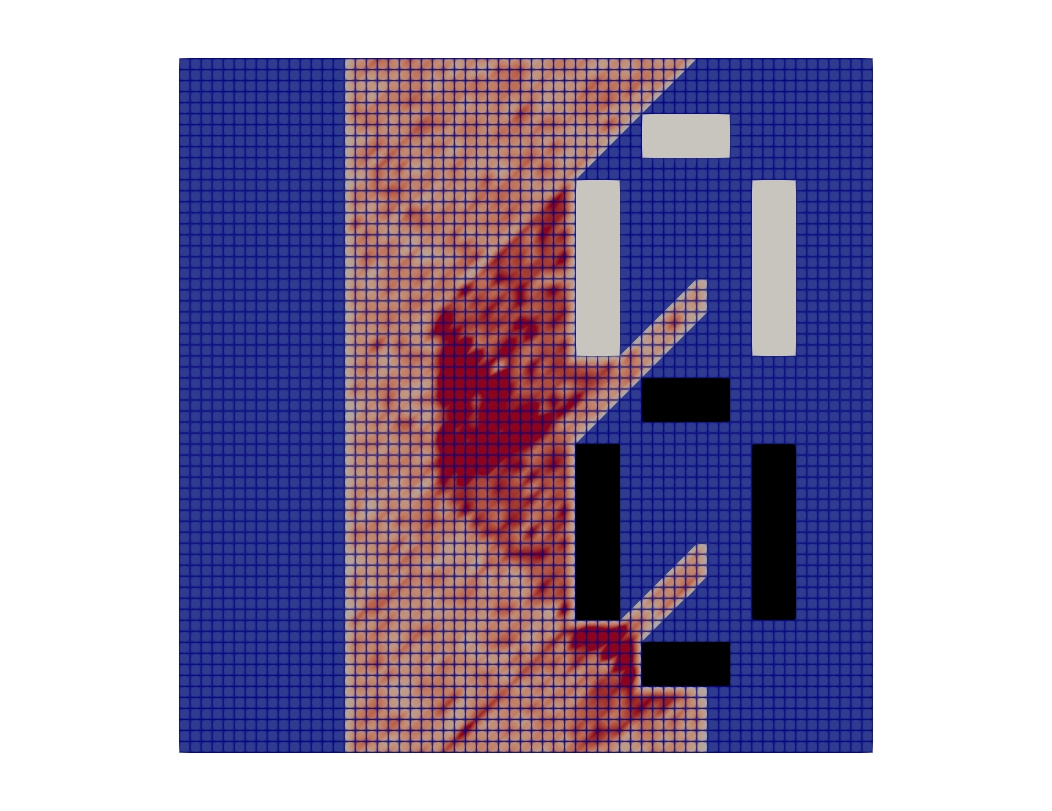}}%
    \\
    \hfill
    \subcaptionbox{The system after 32 time steps.\label{fig:res-4-eight-32}}{\includegraphics[scale=0.22]{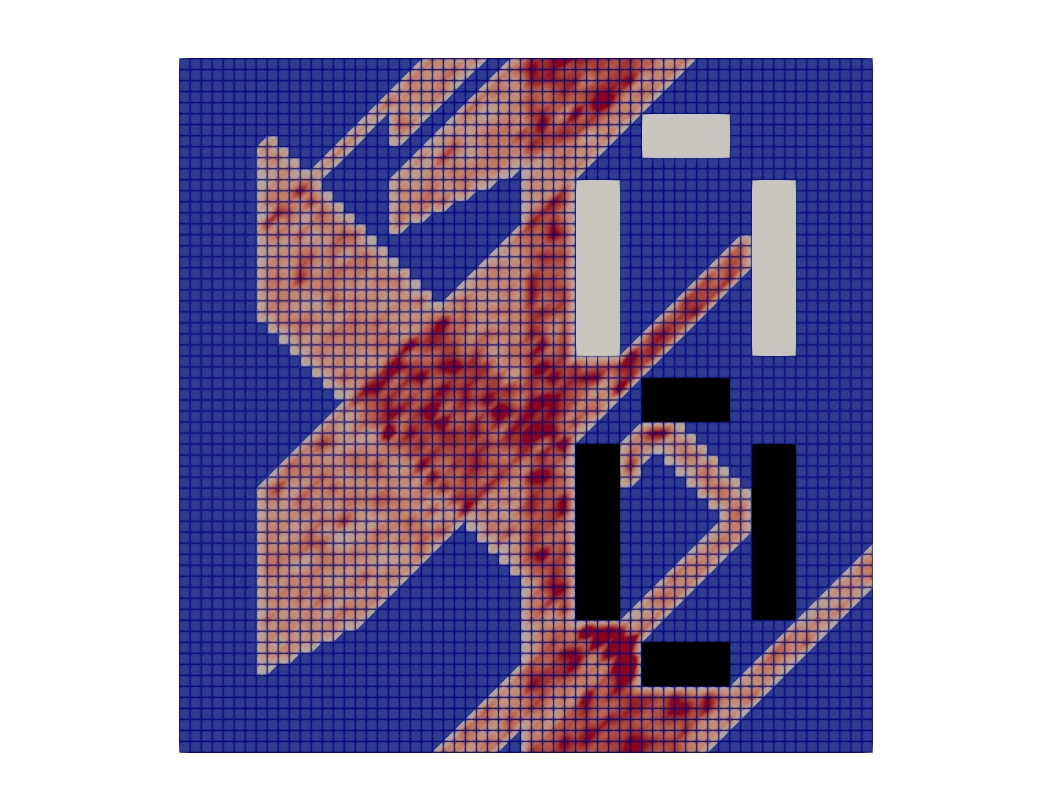}}%
    \hfill%
    \subcaptionbox{The system after 64 time steps.\label{fig:res-4-eight-64}}{\includegraphics[scale=0.22]{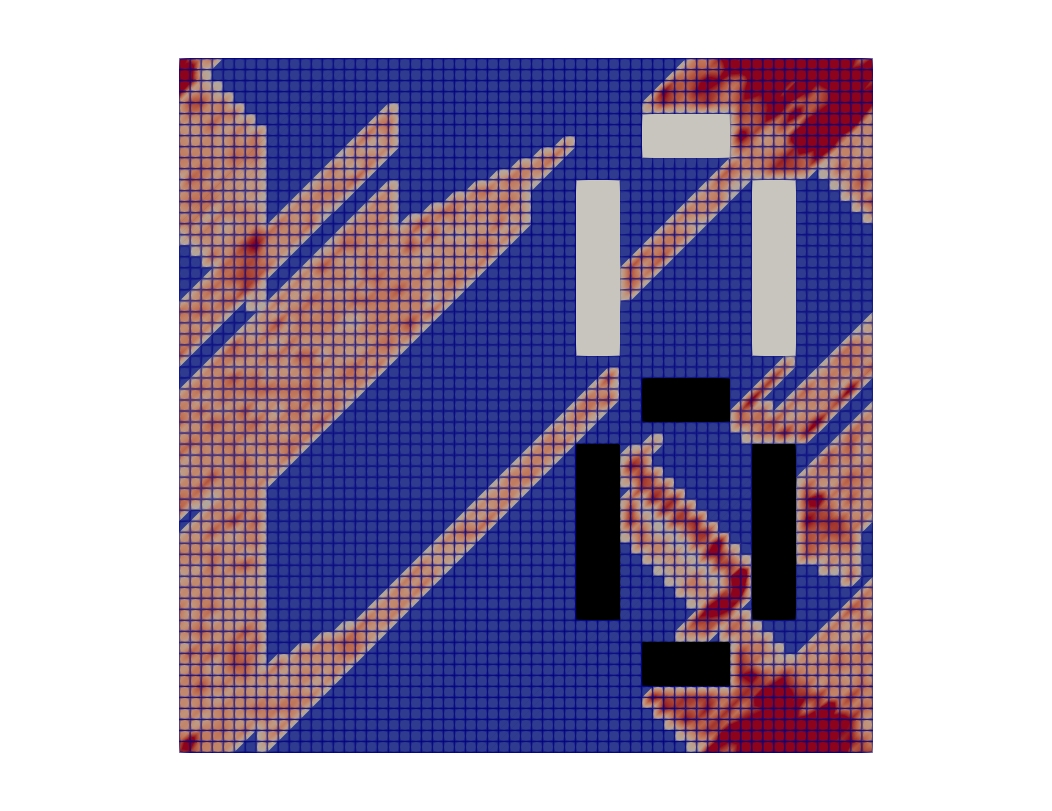}}%
    \caption{Simulation of the QTM algorithm \cite{schalkers2024efficient} on a $64 \times 64$ grid for with 7 solid obstacles for $64$ time steps.}
    \label{fig:res-4-eight}
\end{figure}

\Cref{fig:res-4-eight} depicts the evolution of a 2D system
with 64 grid points in each dimension and seven obstacles placed in
close proximity to one another.
Each dimension has 4 discrete velocities,
and the entire circuit is comprised of only 22 qubits.
Obstacles depicted in grey are imposed bounce-back boundary conditions,
while black objects implement specular reflection.
The edges of the domain implement periodic boundary conditions.
\Cref{fig:res-4-eight-00} shows the initial conditions of the system,
with particles distributed uniformly throughout the left half of the domain.
Darker shades of red indicate the presence of a higher concentration of particles.
Any irregularities in the density stem from the stochasticity
of the counts extracted from the quantum state at the end of each time step.
While inexact, this is the same process that one would follow on actual quantum hardware.
The initial conditions are set up with native quantum gates,
and are such that all particles in the systems have velocities
pointing in the positive directions in both the $x$ and the $y$ axes.
Intuitively, particles are moving towards the upper right-hand corner of the domain.
\Cref{fig:res-4-eight-16}, \Cref{fig:res-4-eight-32}, and \Cref{fig:res-4-eight-64} show
the evolution of the system after $16$, $32$, and $64$ steps.
Higher particle densities emerge naturally at the boundaries of objects,
as well as in areas where particles meet as a result of the change in direction
caused by the different boundary conditions of the objects.
\Cref{fig:res-4-eight-32} showcases the difference between the two types
of boundary conditions: the particles interacting with the obstacles
in the upper half of the domain get reflected along their previous trajectory,
while their conterparts in the lower half of the domain interact differently 
with the obstacle walls.

\begin{figure}
    \centering
    \hfill
    \subcaptionbox{The initial conditions of the particle distribution. \label{fig:res-4-3d-00}}{\includegraphics[scale=0.22]{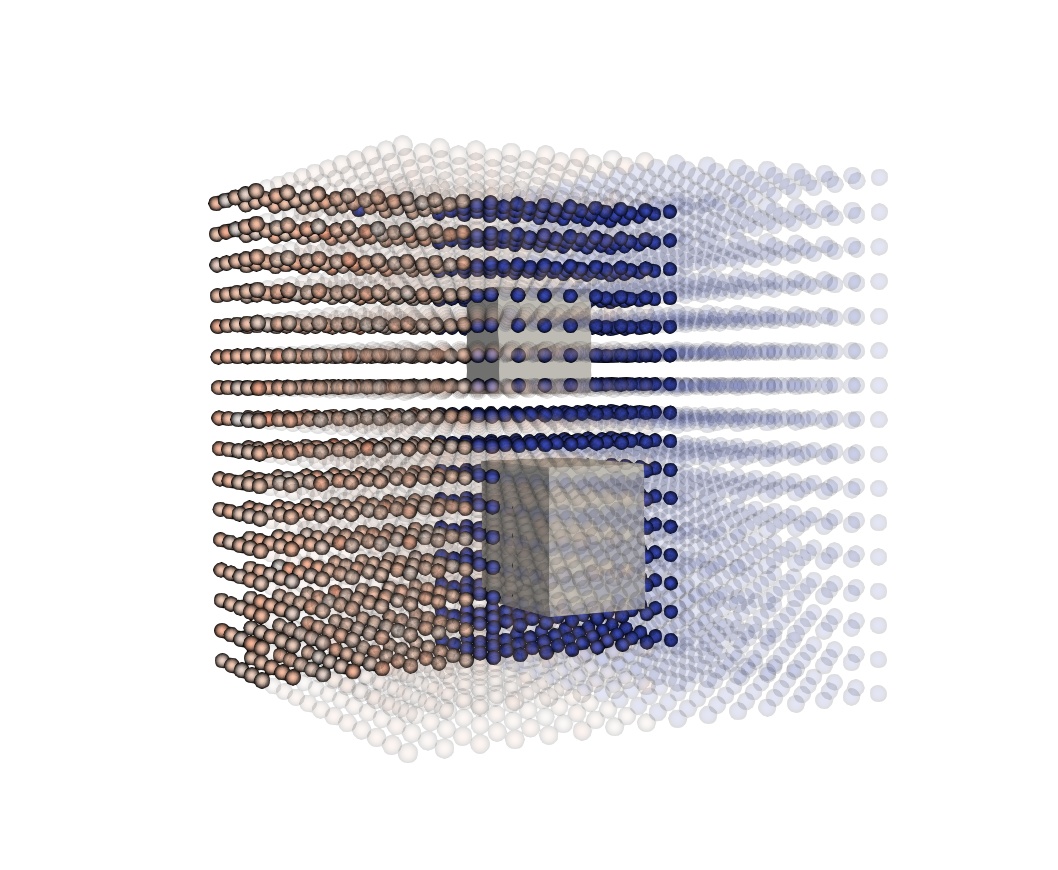}}%
    \hfill%
    \subcaptionbox{The system after 3 time steps.\label{fig:res-4-3d-16}}{\includegraphics[scale=0.22]{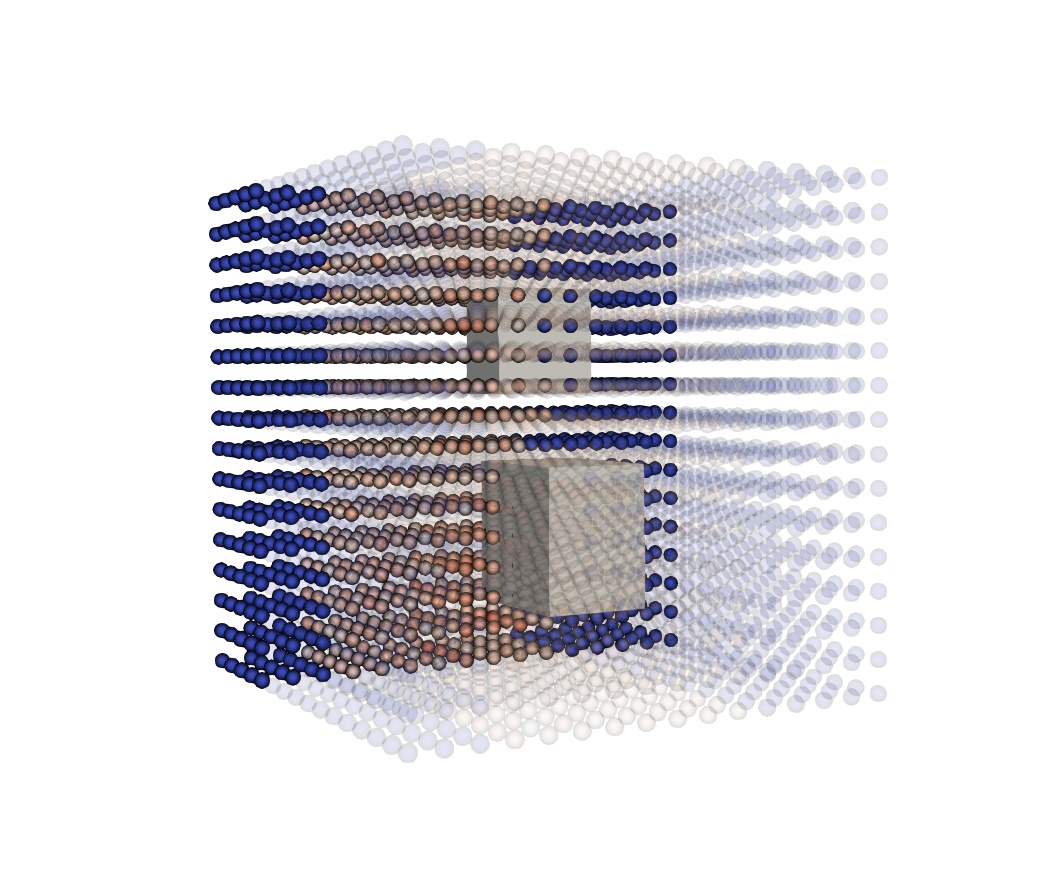}}%
    \\
    \hfill
    \subcaptionbox{The system after 6 time steps.\label{fig:res-4-3d-32}}{\includegraphics[scale=0.22]{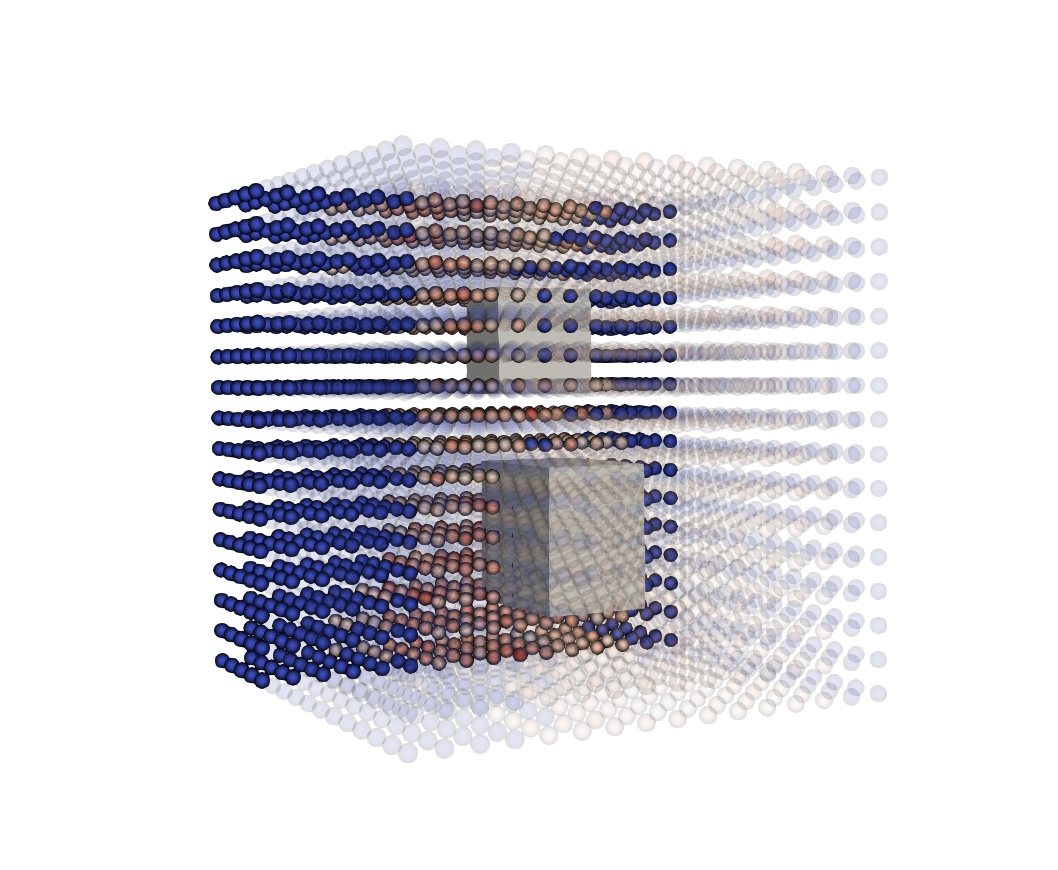}}%
    \hfill%
    \subcaptionbox{The system after 9 time steps.\label{fig:res-4-3d-64}}{\includegraphics[scale=0.22]{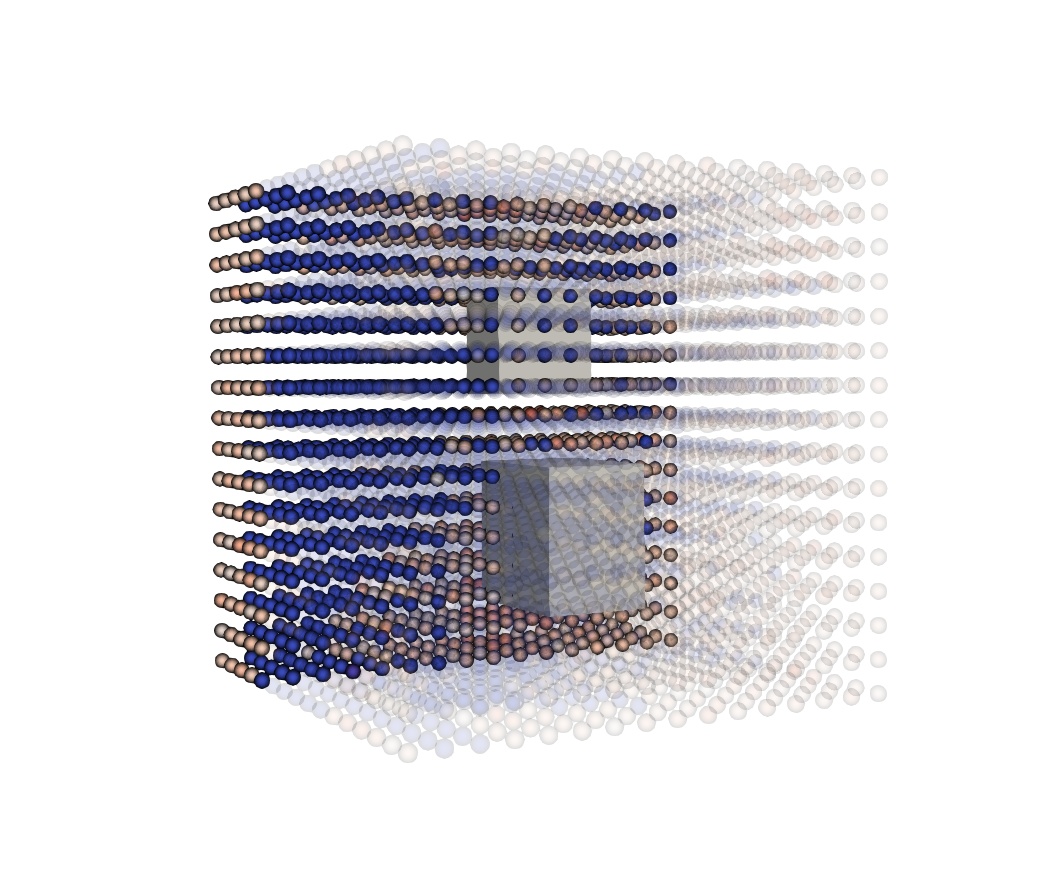}}%
    \caption{Simulation of the QTM algorithm \cite{schalkers2024efficient} on a $16 \times 16 \times 16$ grid for with 2 solid obstacles for $9$ time steps.}
    \label{fig:res-4-3d}
\end{figure}

\Cref{fig:res-4-3d} highlights an example of a 3D
flow in a $16 \times 16 \times 16$ system with
2 bounce-back boundary conditioned obstacles
of different shapes.
As in the previous example, each dimension has 4 discrete velocities.
With \qlbm's adaptive register setup, the entire quantum circuit
only requires 28 qubits.
Each discrete grid point is represented by a sphere, the color of which denotes
the relative density of particles at that physical location.
As in the 2D example, darker shades of red
indicate higher densities, and dark blue spheres
indicate the absence of particles.
\Cref{fig:res-4-3d-00} again shows the initial conditions, which
are the 3D equivalent of the previous example.
We choose this specific visualization integration
as it allows for the examination of individual grid locations
that correspond to specific edge cases in the underlying quantum
algorithm, which makes verification significantly less tedious
than otherwise parsing information from the computed quantum state.
We also highlight the fact that in the \qlbm~implementation
of the QTM algorithm, there are no additional constraints
on 3D systems: the same boundary conditions,
simulation techniques, and visualization media are supported.

\begin{figure}
    \centering
    \hfill
    \subcaptionbox{The initial conditions of the particle distribution. \label{fig:res-4-st-00}}{\includegraphics[scale=0.22]{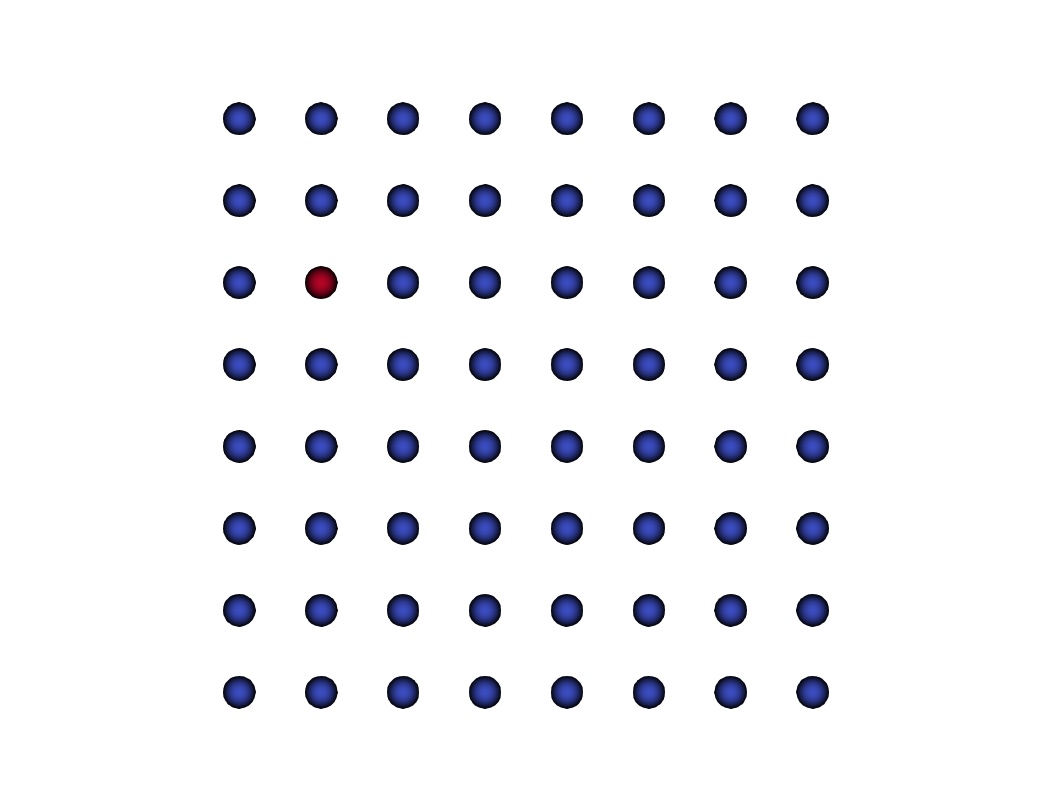}}%
    \hfill%
    \subcaptionbox{The system after 1 time step.\label{fig:res-4-st-01}}{\includegraphics[scale=0.22]{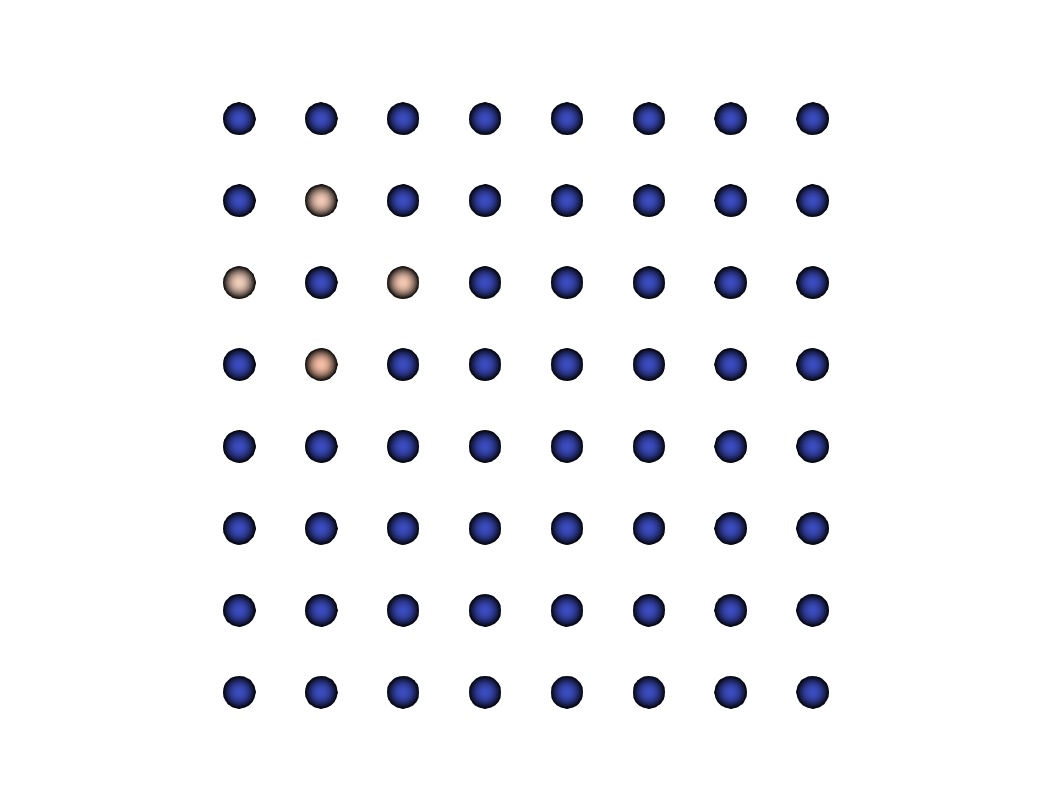}}%
    \\
    \hfill
    \subcaptionbox{The system after 2 time steps.\label{fig:res-4-st-02}}{\includegraphics[scale=0.22]{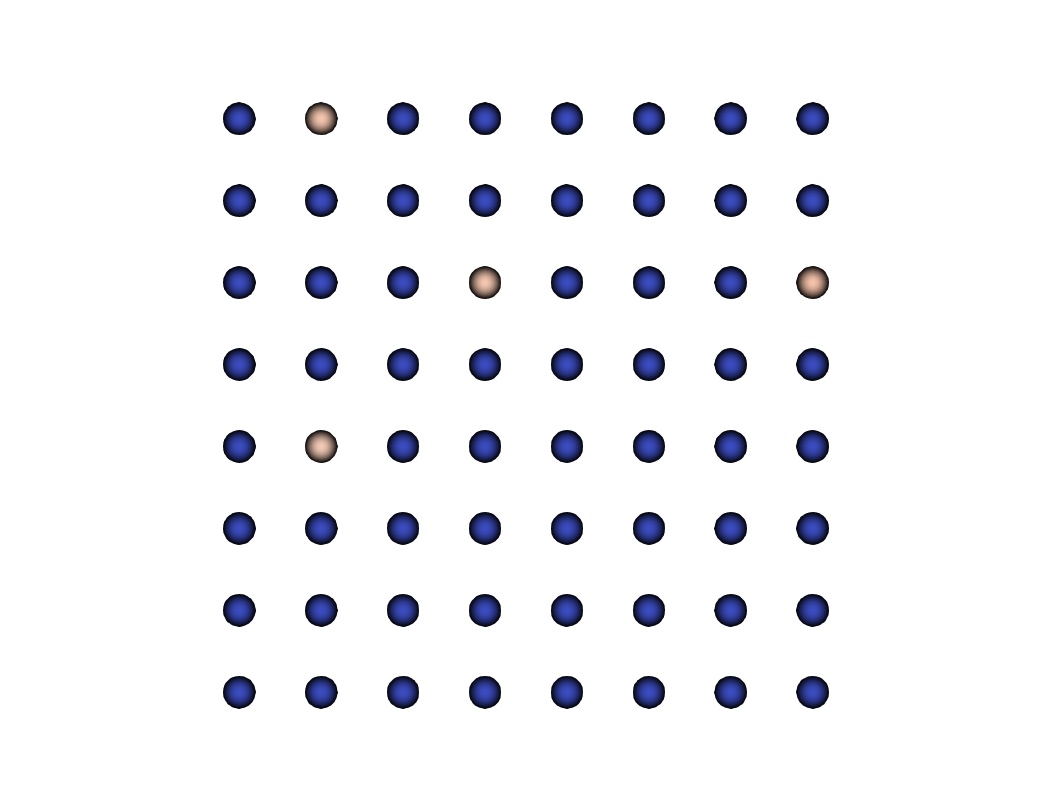}}%
    \hfill%
    \subcaptionbox{The system after 3 time steps.\label{fig:res-4-st-03}}{\includegraphics[scale=0.22]{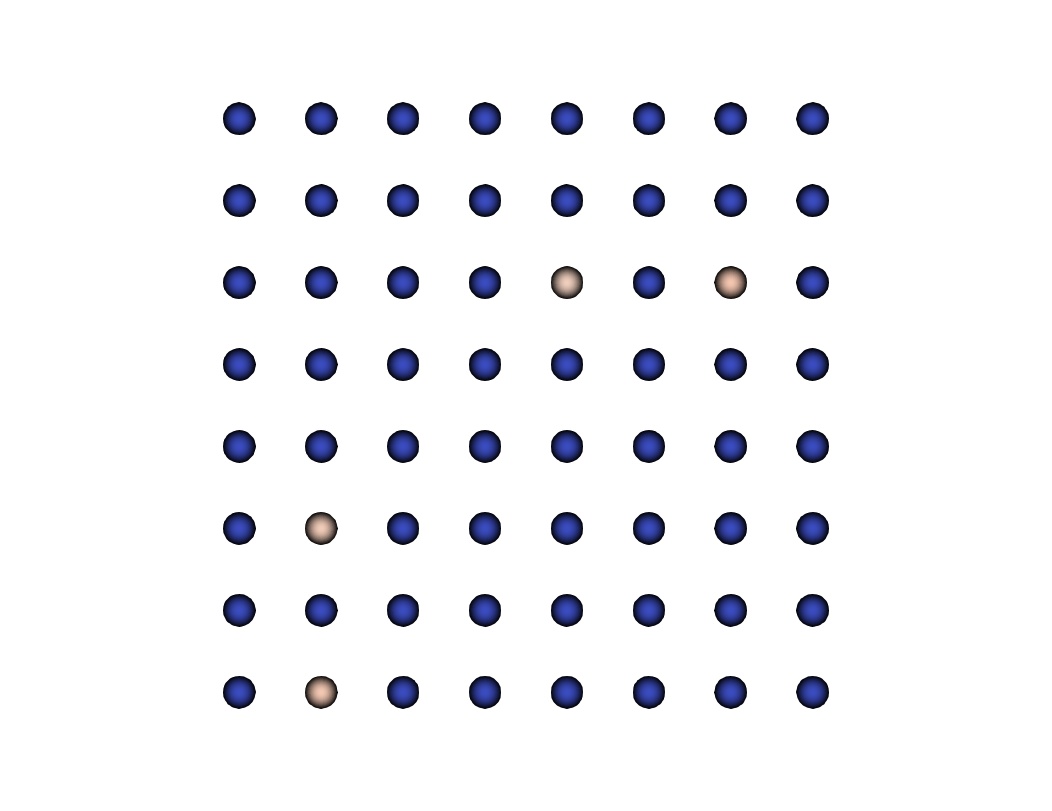}}%
    \\
    \hfill
    \subcaptionbox{The system after 4 time steps.\label{fig:res-4-st-04}}{\includegraphics[scale=0.22]{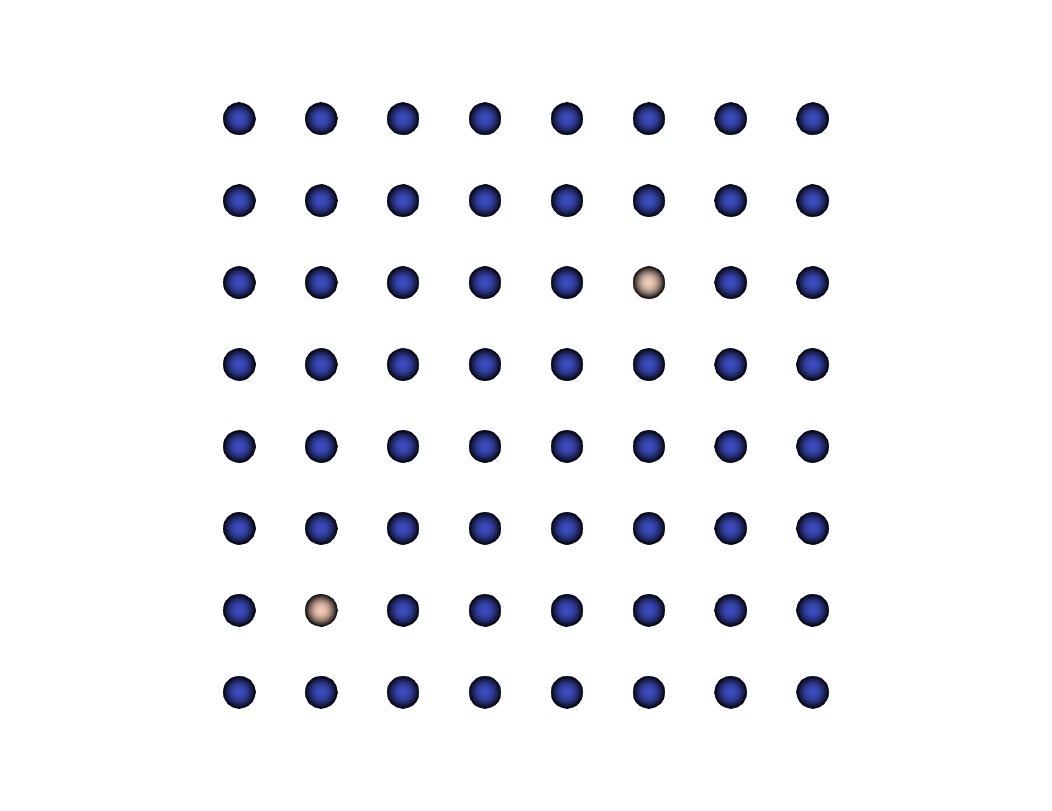}}%
    \hfill%
    \subcaptionbox{The system after 5 time steps.\label{fig:res-4-st-05}}{\includegraphics[scale=0.22]{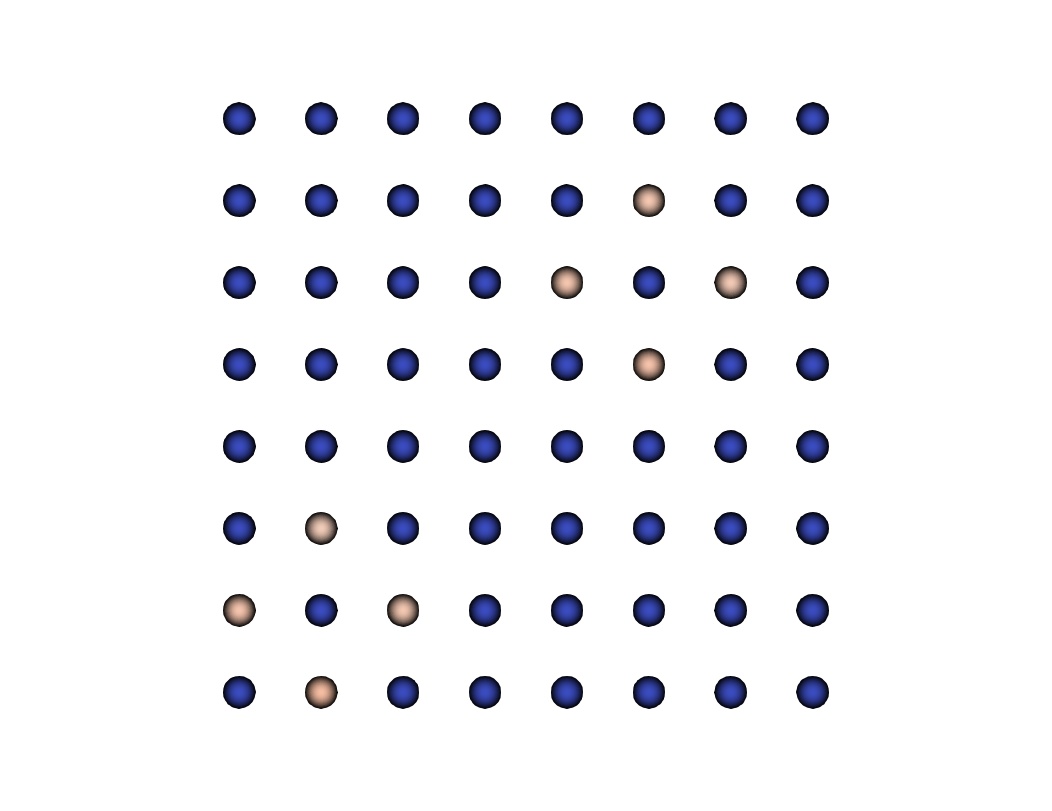}}%
    \caption{Simulation of the STQBM algorithm \cite{schalkers2024importance} on a $16 \times 16 \times 16$ grid for with 2 solid obstacles for $64$ time steps.}
    \label{fig:res-4-st}
\end{figure}

\Cref{fig:res-4-st} illustrates the evolution of a $8 \times 8$
 system simulated with the STQBM algorithm.
In addition to streaming, the STQBM also performs collision 
at the cost of including neighboring velocity information for each grid point.
In practice this limits the size of systems that classical hardware
can emulate for practical development and research purposes.
For the 5 steps in \Cref{fig:res-4-st}, $1024$ qubits would have been required
to simulate the end-to-end system, which is infeasible
for any classical hardware available today.
The simulation was instead performed using \qlbm's automated reinitialization
mechanism described in \Cref{subsec:computational-imporvements}, which
can function with circuits as small as one time step.
While this does inherently introduce inaccuracies in the
quantum computation, the space-time encoding 
is less susceptible to this than amplitude-based methods,
and the performance advantage gained from reinitialization is substantial
-- the entire simulation, including parsing the results into
the visualization format takes seconds on commodity hardware.
The only precision lost through reinitialization is
in the relative density of particles at specific grid locations --
basic constraint such as conservation of mass are not violated. 
We again stress that reinitialization is a feature of \qlbm~and
not a requirement of the underlying algorithms,
which can be executed entirely on quantum hardware,
provided a sufficient number qubits.

\begin{figure}
    \centering
    \includegraphics[scale=0.5]{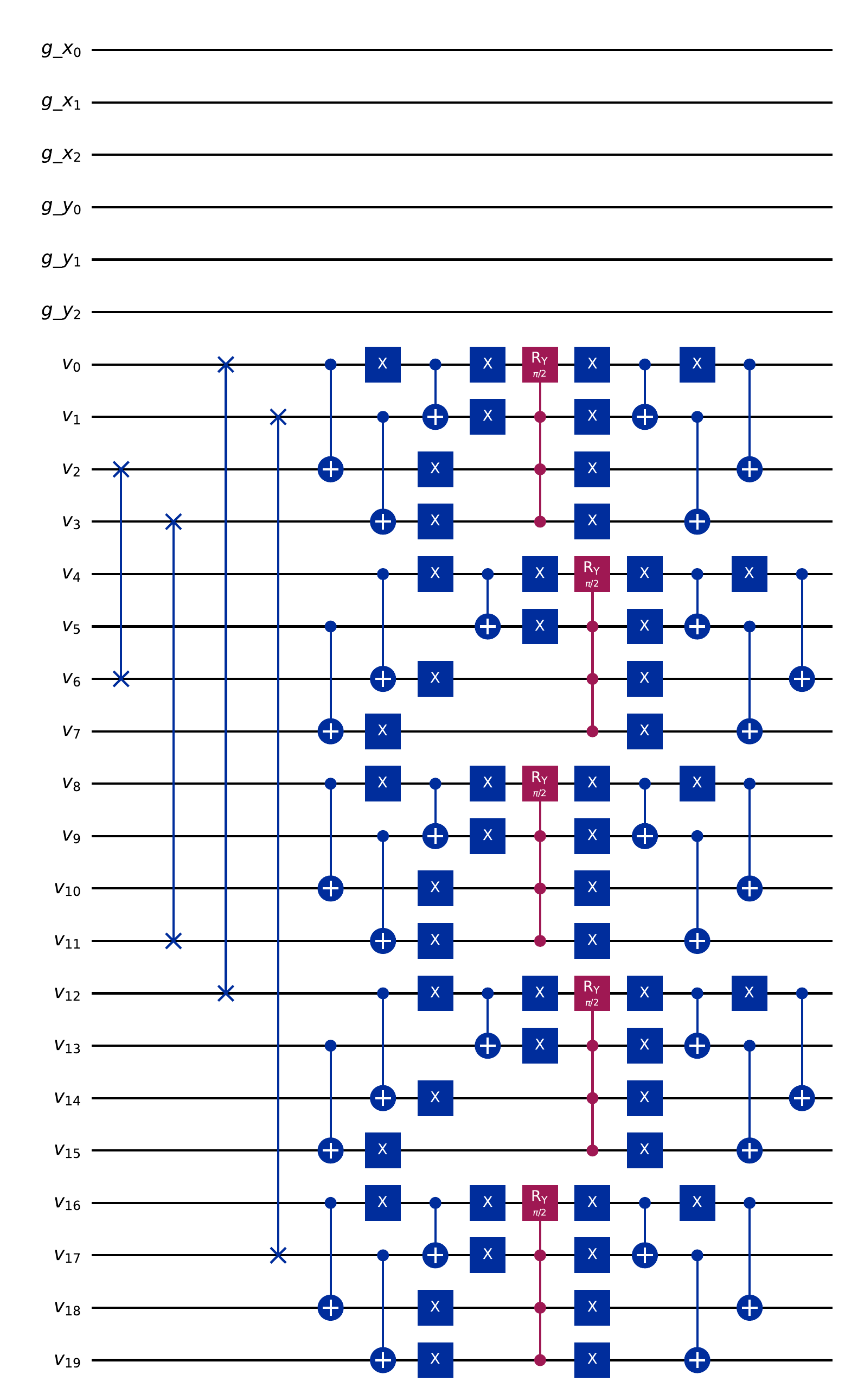}
    \caption{One time step STQBM \cite{schalkers2024importance} circuit.}
    \label{fig:res-4-st-circuit}
\end{figure}

\Cref{fig:res-4-st-00} shows the initial state of the system, where
4 particles are concentrated in one grid point, under the $D_2Q_4$
lattice discretization.
The 4 particles are each travelling along
one of each of the $4$ discrete velocity channels.
Following one time step, particles stream to neighboring grid points (\Cref{fig:res-4-st-01}).
Since collision only affects instances where particles reach
the same grid point and have a velocity profile that can be mapped onto an
equivalence class, the $3$ following time steps are unaffected by it
and only practically consist of streaming, with the added complexity of periodicity.
In \Cref{fig:res-4-st-04}, two particles reach the same grid point
in two different instances, both with velocity profiles that collision affects.
As a result, collision redistributes the particles such that mass and momentum
are conserved, as described in \cite{schalkers2024importance}.
This simulation was carried out using a circuit that only performs 
the computation of one time step, shown in \Cref{fig:res-4-st-circuit}.
Reinitialization automatically computes the initial conditions
that allow the transition between steps be carried between time steps.
This is assumed to prepare the state of grid qubits prior to simulation.

\section{Conclusion\label{sec:conclusion}}

We introduced \qlbm, a Python software framework
that aims to accelerate the development, simulation,
and analysis of Quantum Lattice Boltzmann Methods.
We designed \qlbm~as an end-to-end development environment
that caters to every step of the research process,
from assembling proof-of-concept quantum circuits
to analyzing their performance within different simulation platforms.
The modular architecture of \qlbm~decouples the hierarchically
arranged quantum component module from external infrastructure,
which promotes testability through isolation.
Additional modules interface with state-of-the-art
quantum simulators and compilers, which allows users
to seamlessly tune their setup according to their goals and resources.

To increase the accessibility of QBMs to researchers and practitioners
with various backgrounds, we implemented convenient interfaces
that bridge the gap between the delicate quantum circuit assembly process
and the high-level interfaces that users have come to expect
from more mature classical software frameworks.
We introduced novel simulation techniques in the form of statevector
snapshots, statevector sampling, and reinitialization,
which massively increase the
performance of simulations and in turn hasten future research.
Finally, we demonstrated the versatility of these techniques by incorporating
them within 2D and 3D simulations on CPUs and GPUs
and showed the practical benefit of
built-in experimentation pipelines and visualization techniques.
To encourage collaboration and reproducibility in the field
of quantum computational fluid dynamics, we make both
the source code of \qlbm~and a replication package of this study available
at \url{https://github.com/QCFD-Lab/qlbm} and 
\cite{georgescu_2024_14231193}, respectively.

We envisage two future directions for \qlbm.
Primarily, the purpose of this framework is
to be of service the broader QCFD research community
and assist in QBM development.
This includes both the implementation of novel
algorithms in the future, as well as the generalization
of past and present techniques from the literature.
Secondarily, \qlbm~will remain up-to-date
with the rapid developments occurring in the quantum software field.
Novel simulation and transpiler technologies, and access to increasingly 
robust quantum hardware are developments which we aim to continue
to integrate within \qlbm,
while retaining its high code quality and reproducibility standards.
\section{Acknowledgements\label{sec:ack}}

We gratefully acknowledge support from the joint research program
\emph{Quantum Computational Fluid Dynamics} by Fujitsu Limited and Delft
University of Technology, co-funded by the Netherlands Enterprise
Agency under project number PPS23-3-03596728.


\bibliographystyle{plainnat}
\bibliography{references}
\end{document}